\documentclass[12pt,letterpaper]{article}

\usepackage[margin=1in]{geometry}
\usepackage{newtxtext,newtxmath}
\usepackage{authblk}

\usepackage{natbib}
\usepackage{subcaption}
\usepackage{multirow}
\usepackage{tikz}
\usepackage{csquotes}
\usepackage{enumerate}
\usepackage{hyperref}
\usepackage{array}
\usepackage{booktabs}

\usetikzlibrary{shapes.geometric, arrows}

\tikzstyle{startstop} = [rectangle, rounded corners, 
minimum width=3cm, 
minimum height=1cm,
text centered, 
draw=black]

\tikzstyle{io} = [trapezium, 
trapezium stretches=true,
trapezium left angle=70, 
trapezium right angle=110, 
minimum width=3cm, 
minimum height=1cm, 
text centered, 
draw=black]

\tikzstyle{process1} = [rectangle, 
minimum width=3cm, 
minimum height=1cm, 
text centered, 
text width=6cm, 
draw=black]

\tikzstyle{decision1} = [diamond, 
minimum width=3cm, 
minimum height=1cm, 
text centered, 
text width=2.5cm,
draw=black]

\tikzstyle{decision2} = [diamond, 
minimum width=0.5cm, 
minimum height=0.5cm, 
text centered, 
text width=3.0cm,
draw=black]

\tikzstyle{arrow} = [thick,->,>=stealth]

\title{Robust Parameter Estimation for Snow Load Induced by Annual Maximum Snow Accumulation Using Constrained Bayesian Priors}

\author{
Shaveen A. Britto$^{1}$ \quad and \quad Brennan L. Bean$^{1}$ \\
{\small $^{1}$Department of Mathematics and Statistics, Utah State University, Utah, USA}
}

\date{}

\begin{document}
\maketitle

\begin{abstract}

This paper develops a Bayesian framework to estimate the parameters of the Generalized Extreme Value (GEV) distribution for the weight induced by annual maximum accumulations of snow, referred to as the snow load, using a Hamiltonian Monte Carlo (HMC) algorithm as implemented in the \enquote{extremMHMC} R package developed alongside this paper. Key to the approach is the use of strong prior distributions for the shape parameter that are appropriate in the context of snow loads, which helps to ensure robustness in the distribution parameter estimates for annual maximum snow loads despite small sample sizes. This robustness is key to ensuring that structural reliability analyses, which rely on the GEV distribution, produce physically realistic estimates of snow loads. Information on strong prior distributions is derived from existing studies on extreme rainfall and snowfall, with the novel use of hyperbolic tangent functions to transition the prior distribution parameters between low and high snow regimes. This approach enables global applicability of the strong prior approach while maintaining physical realism. Additionally, simulation studies confirm a reduction in Root Mean Square Error (RMSE) of the shape parameter estimate compared to frequentist methods, particularly for small sample sizes. Finally, a real-world application using a data from 9715 stations further demonstrates the feasibility of the proposed Bayesian framework for large scale implementation.

\end{abstract}


\maketitle

\renewcommand\thefootnote{}

\renewcommand\thefootnote{\fnsymbol{footnote}}
\setcounter{footnote}{1}

\section{Introduction}\label{section:introduction}

Many studies in Climate Science related to Extreme Value Theory (EVT) focus on extreme events in Rainfall \citep{Papalexiou2013battle}, Heatwaves \citep{Philip2022rapid}, Flooding \citep{Mushtaq2022reliable}, Wind Patterns \citep{Sarkar2019weibull}, and Hydrology \citep{Martins2000generalized}. However, minimal studies have been conducted on EVT for the weight induced by extreme snow accumulations, referred to hereafter as the snow load. Extreme snow loads can create significant impacts in various fields, including agriculture \citep{Vico2014snowed}, energy production \citep{Dahlioui2025snow}, transportation \citep{Li2025snow}, and particularly in structural engineering and construction \citep{Geis2012Snow}. 

Historically, "design" snow loads were based on a 50-year mean recurrence interval (MRI), also known as the uniform hazard approach. These 50-year MRI snow loads were then multiplied by a safety factor that was calibrated to achieve a desired level of safety in the average \citep{Ellingwood1980development}. However, this approach results in differing levels of safety based on local snow accumulation dynamics, as was highlighted in the state of Colorado \citep{DeBock2016colorado}. Recently, engineering codes have shifted towards a site-specific reliability-targeted snow loads (RTSL) approach, also known as the uniform risk approach \citep{Bean2021SnowLoad}. The RTSL approach simultaneously considers the uncertainty in the environmental hazard and the structural design to prescribe a design load that is calibrated to ensure an acceptably low probability of structure failure. In both the uniform hazard and uniform risk approach to design, the most critical step is fitting a probability distribution that appropriately describes the extreme right tail of annual maximum ground snow loads. Currently, the distribution of choice for fitting annual maximum ground snow loads is the Generalized Extreme Value (GEV) distribution, as adopted in ASCE/SEI 7-22 \citep{ASCE722}.

The GEV distribution is a three-parameter distribution characterized by location ($\mu$), scale ($\sigma$), and shape ($\xi$) parameters. Accurate estimates of these three parameters are crucial for the structural reliability analysis, especially the shape parameter due to its substantial influence on the extreme quantile estimates of the distribution. Various studies have been conducted to estimate the shape parameter in the field of climate science using both Frequentist and Bayesian methods. However, in the context of snow load hazards, there have been no studies on estimating parameters with the Bayesian methods and only few studies have been conducted with the Frequentist methods.

To address this gap, this paper focuses on estimating GEV parameters using a Bayesian framework that incorporates domain knowledge about extreme rainfall and snow load events into strong prior distributions. This Bayesian framework offers a substantial advantage in estimating parameters by allowing prior information to complement limited sample sizes. The result is a robust Bayesian alternative to extreme snow load probability distribution fitting that is demonstrated to be feasible for use on a network of 9000+ snow measurement locations. 

The remainder of this paper proceeds as follows: We first present the background related to the GEV distribution, Hamiltonian Monte Carlo (HMC) algorithm, and prior distributions for shape parameter in Section \ref{section:background}. In Section \ref{section:methodology}, we summarize the methodology with the Bayesian formulation, selecting a prior distribution for the shape parameter and estimating RTSL, followed by two simulation studies in Section \ref{section:simulation_study} and application for a real-world dataset in Section \ref{section:application}. Finally, Section \ref{section:conclusion} presents the overall conclusions of the study and discusses the implications of our results.

\section{Background}\label{section:background}

The use of the GEV distribution for extreme events is based on Fisher-Tippett-Gnedenko theorem \citep{FisherTippett1928limiting, Gnedenko1943sur}, which states that the distribution of properly normalized block maxima converges to one of three families of distributions, including Gumbel, Fr\'echet, and Weibull. Later \cite{Jenkinson1955thefrequency} proposed the GEV distribution, which encompasses the above three distributions into a single distribution through the incorporation of a third distribution parameter called the shape. The GEV probability density function is represented mathematically as, 
\begin{equation}\label{eq:gevpdf}
\begin{aligned}
f_X(x;\mu,\sigma,\xi) = \begin{cases}
\frac{1}{\sigma} \left(1 + \xi \left( \frac{x - \mu}{\sigma} \right) \right)^{-\left(1+\frac{1}{\xi} \right)}\exp \left\{ - \left( 1 + \xi \left( \frac{x - \mu}{\sigma} \right) \right) ^{-\frac{1}{\xi}} \right\} &, \xi \neq 0 \\
\\
\frac{1}{\sigma} \exp\left\{ - \left( \frac{x - \mu}{\sigma} \right) - \exp\left(-\frac{x - \mu}{\sigma} \right)\right\} &, \xi = 0
\end{cases}    
\end{aligned}    
\end{equation}
where $\mu \in \mathbb{R}, \ \sigma > 0, \ \xi \in \mathbb{R}$. 

The Fr\'echet, Weibull, and Gumbel families are obtained for $\xi > 0$, $\xi < 0$, and $\xi = 0$, respectively. The Gumbel distribution ($\xi = 0$) has an exponentially tailed distribution, with a rapidly decreasing tail; the Weibull distribution ($\xi < 0$) has a light tail finite upper bound in the support, and the Fr\'echet distribution ($\xi > 0$) is heavy-tailed, with a slowly decreasing tail \citep{Bali2003generalized}. 

\subsection{Bayesian Framework} 

In this paper, we focus on the GEV distribution for estimating extreme snow loads using a Bayesian framework. Bayesian approaches provide a robust framework for integrating prior knowledge with observed data, particularly in situations where extreme events are of interest or where data are scarce. We select a simulation-based Hamiltonian Monte Carlo (HMC) algorithm as an efficient sampling framework for Bayesian inference, since the GEV likelihood combined with physically motivated priors yields non-conjugate posterior distributions. HMC is particularly well-suited for this due to its efficiency in exploring high-dimensional and correlated parameter spaces even when working with small sample sizes, which is typical in extreme value applications. The resulting posterior samples enable direct quantification of uncertainty in the GEV parameters.

HMC employs equations that describe particle motion in space, known as Hamiltonian Dynamics, which assumes no friction in a physical analogy. For the Bayesian framework, HMC is an MCMC sampler designed to efficiently explore complex posterior distributions by leveraging gradient information to propose new samples. The core concept of this algorithm is that particles move over a surface without any energy loss due to friction. HMC was initially introduced by \cite{Duane1987hybrid} under the name "Hybrid Monte Carlo". While HMC uses the Metropolis-Hastings (MH) framework \citep{Metropolis1953equation, Hastings1970monte}, it has an advantage compared to traditional MH because the “particle” travels along trajectories guided by the gradient of the log-posterior, allowing it to traverse high-density regions and avoid the random-walk behavior of traditional MH, and this leads to obtaining higher and balanced acceptance rates for proposed samples compared to MH. Overall, HMC is considered a method for generating effective proposals for target distributions.

In Bayesian statistics, the MH algorithm tends to generate new samples from regions of the posterior distribution with high posterior density. In contrast, the HMC algorithm explores regions by introducing auxiliary momentum variables and uses the negative log-posterior (which, in statistical mechanics, is analogous to potential energy) together with its gradient, to simulate Hamiltonian dynamics in an extended state space to generate new samples from the posterior distribution \citep{Neal2011mcmc, Betancourt2017aconceptual}. Apart from the standard construction of HMC, which has a negative log-posterior as the potential energy, using a negative log-posterior has some practical advantages. Negative log-posterior improves the numerical stability and the gradient scaling in HMC and this leads to more efficient proposals, reduced random-walk behavior, and better exploration of complex non-conjugate posterior distributions. For more details regarding the background and methodology of the HMC algorithm, please refer to the appendix (see Appendix~\ref{appendix:A1}). 

GEV Likelihood function combined with a geophysically motivated prior usually produces a posterior distribution with a complex geometry (meaning it can be difficult to sample from) and strong parameter dependence between parameters, especially involving the $\xi$ parameter. In such cases HMC offers the following substantial advantages for generating efficient samples from the posterior distribution compared to the MH algorithm:

\begin{itemize}
    \item \textbf{Guide the sampling process with gradients:} Make significant moves in parameter space while still being likely to be accepted. It utilizes gradient information to navigate parameter space effectively. These gradient informed trajectories reduce random-walk behavior and improve mixing compared to the traditional MH algorithm \citep{Neal2011mcmc, Betancourt2017aconceptual}.
    \item \textbf{Reduced correlation between generated samples:} HMC uses simulated dynamics to propose moves, which means that the samples generated are less correlated. This means a few samples is enough to estimate moments of the target distribution. This reduces computation time and saves computational resources \citep{Neal2011mcmc, Betancourt2017aconceptual}.
    \item \textbf{Reliable tail exploration:} HMC algorithm predominantly samples from regions of higher posterior density, but still effectively samples the tail regions of the distribution as well. Sampling these tail regions is crucial in our study because the accurate estimation of the shape parameter in the GEV distribution depends on effectively capturing the tail behavior \citep{Betancourt2017aconceptual}. 
\end{itemize} 

These advantages are more important when estimating $\xi$, whose influence dominates the super-extreme ground snow load estimates that govern design provisions. To reduce manual hyperparameter tuning and improve computational efficiency, we incorporate the dual-averaging algorithm introduced by \citet{Hoffman2014thenuts} in our HMC implementation. 

The application of Hamiltonian Monte Carlo (HMC) to Bayesian inference in GEV models was examined by \citet{Hartmann2017bayesian}, which used Normal distribution priors to compare it with the traditional Metropolis-Hastings (MH) algorithm. They applied normal priors to the transformed parameters and used HMC with a step size of 0.12 and 27 leapfrog steps using annual maximum sea level data from Port Pirie (1923-1987). According to their findings, HMC generated samples with a higher acceptance rate (approximately 0.95) and significantly lower autocorrelation than MH. Significantly larger effective sample sizes for HMC were suggested to result in less sample dependence. Their study showed that, with sample sizes $n = 15, 30, 50, 100$ and GEV parameters $\mu = 2$, $\sigma = 0.5$, and $\xi = -0.1$, HMC produced parameter estimates with Mean Square Error (MSE) than MH in simulation studies.

\subsection{Prior for the Shape Parameter in Extreme Snow Loads}

For practical purposes, climate scientists bound the $\xi$ parameter range between $-0.5$ and $0.5$, as the GEV distribution has infinite variance outside of that range \citep{Coles2001anintroduction}. However, there are additional practical limitations for tail extrapolation when $|\xi| > 0.25$ due to the impractical tail extrapolations that occur for super-extreme percentiles. Unfortunately, the large natural variability in estimates of $\xi$ with small sample sizes causes estimates of $\xi$ to regularly fall outside the practical bounds. This motivates the need for strong prior information that will reduce the variability of the estimates of $\xi$. For those strong priors, we look to information related to the distribution of all forms of precipitation (i.e., snow and rain) because snow is fundamentally a precipitation-driven process. Our consideration of rainfall distributions is motivated by the different ways in which extreme snow accumulates. In warm climates, air temperatures are typically above freezing and the annual extreme snow event is often produced by only one or two snow storms, since snow that does accumulate usually melts quickly. On the other hand, in colder climates, snow accumulates for long periods of time as an accumulation across many different storms. In cold climates, we expect that distributions for extreme rainfall will have heavier tails than a season long accumulation, due to the normalization effect that occurs when accumulating several extreme precipitation events over a long period of time. 
 

To motivate prior distributions for the GEV shape parameter, we review existing studies that estimate $\xi$ for precipitation and rainfall-related extremes. Table~\ref{tab:shapelit} summarizes the measurement type, data source, sample size, and reported values of $\xi$ across this literature. Together, these studies provide empirical guidance on plausible ranges for $\xi$ under different climatic and methodological settings.

\cite{Shamseldin2010generalized} used station-level point observations from NCDC (National Climatic Data Center), along with gridded data from NCEP (National Centers for Environmental Prediction) Reanalysis and the CCSM (Community Climate System Model) climate model, to analyze extreme daily precipitation amounts. They used "point-process approach" with a high percentile threshold to model with GEV instead of "Block Maxima approach". MLE was used to estimate the parameters for both gridded and station series. Gridded data are measurements or climate model outputs mapped onto a regular lattice of equally sized grid cells covering a region, where each cell stores a value representing climate conditions for that area. The main finding was that the GEV shape parameter is typically positive throughout the seasons, indicating heavy-tailed behavior. Across all seasons, the estimated $\xi$ is generally small and tends to be positive on average, but negative estimates also occur. The average $\xi$ for station data varies from approximately 0.087 (spring) to 0.127 (summer). For gridded (NCEP) data, the range is approximately 0.04 (summer) to 0.12 (fall). Despite the fact that the mean estimated shape parameter is positive, a substantial fraction of locations still yield negative estimates of $\xi$: only 64\% and 77\% of station-based estimates are positive in spring and fall, respectively, and only 41\% of grid-cell estimates are positive in spring. This suggests that the tail heaviness is generally consistent across datasets, although the degree of tail heaviness shows a noticeable seasonal and data-source (i.e., station vs gridded) variation. Winter is defined as DJF (December, January, and February), but their study does not explicitly state whether winter precipitation measurements treat snowfall as liquid-water equivalent or some other form. 

To evaluate tail behavior in rainfall extremes, \cite{Papalexiou2013battle} examined the annual maxima of daily rainfall from 15,137 GHCN-Daily records worldwide. The estimated $\xi$ parameters lie between -0.59 and 0.76 using L-moments for all records with mean 0.093 and 90\% empirical confidence interval (ECI) [-0.11, 0.28]. The authors' record-length adjusted (limiting) analysis suggests that the across-station GEV shape parameter is approximately Normal distributed with limiting mean 0.114 and limiting standard deviation 0.045 and under this Normal approximation, the central 95\% and 99\% intervals are approximately $[0.03,\,0.21]$ and $[0.00,\,0.23]$, respectively. They also mentioned that estimated values of $\xi$ primarily fall within a relatively narrow range such as 0 to 0.23 (consistent with predominantly heavy-tailed behavior), indicating moderately heavy but bounded upper tails for extreme daily rainfall across the majority of regions, despite obvious regional variability. 

Extreme precipitation patterns were described by research from \cite{Ragulina2017generalized} using a Bayesian hierarchical GEV model with the Metropolis-Hastings MCMC algorithm. Based on data from 71 Norwegian stations and the GHCN-Daily (Global Historical Climatology Networks), the analysis found that the tail index is generally positive, with a mean posterior estimate of $\xi \approx 0.139$ (with the 95\% credible interval (CrI) of (0.127,0.150)). One of the main conclusions of their study was the substantial negative impact of elevation on $\xi$, which falls by roughly 0.07 for every 1000 meter increase in altitude. This effect causes substantial spatial heterogeneity: high-elevation sites frequently follow a Gumbel or even negative-shape distribution ($\xi \in [-0.03, 0.17]$), while sea level stations exhibit a broad range of tail heaviness ($\xi \in [0.04, 0.24]$). Consequently, both regional variability and the altitudinal damping of precipitation extremes were effectively characterized by the Bayesian framework. 

The GEV analysis of snow is less prevalent than precipitation studies. However, there are some notable studies that provide insight into snow accumulation for high snow regions. \cite{Roux2023projection} fit a non-stationary GEV model to annual maxima of daily snowfall for the 23 French Alpine mountain regions at 300m elevation bands from 900 m to 3600 m with Global Mean Surface Temperature (GMST) used as the lone covariate. They adopt a piecewise-linear model for $\mu(T)$, $\log(\sigma(T))$, and $\xi(T)$ and used MLE to estimate the coefficients. A supplement to this study reported that across elevations and temperature levels, the shape parameter $\xi$ generally lies around -0.10 to +0.17, in one case reaching $\approx 0.40$ at $+4$ $^{\circ}$C, and remains within a physically plausible band of $[-0.5, 0.5]$. \cite{Roux2023projection} results focus on snowfall extremes, which may be related to snow loads because large accumulation events are often driven by intense snowfall.

Table~\ref{tab:shapelit} summarizes these observed ranges for different precipitation forms. This paper uses these studies as an empirical anchor to specify strong prior distributions for the GEV shape parameter, as the reported ranges provide a defensible target for constraining $\xi$ under precipitation extremes. Concretely, we choose baseline priors whose central mass covers the bulk of values in Table~\ref{tab:shapelit}.

\begin{table}[htbp]
\caption{Summary of reported GEV shape parameter ($\xi$) ranges in the literature.}
\label{tab:shapelit}
\centering

{\footnotesize

\begin{tabular*}{\textwidth}{@{\extracolsep{\fill}}c c c c c@{}}
\hline
Study &
\begin{tabular}{@{}c@{}}Climate \\ Variable \end{tabular} &
\begin{tabular}{@{}c@{}}Number of \\ Locations \end{tabular} &
\begin{tabular}{@{}c@{}}Number of \\ years or \\ Year range \end{tabular} &
\begin{tabular}{@{}c@{}}Typical Range \\ for $\xi$ \end{tabular} \\
\hline

\cite{Shamseldin2010generalized} &
\begin{tabular}{@{}c@{}}Daily \\ precipitation \\ extremes \end{tabular} &
\begin{tabular}{@{}c@{}}$\approx 4300+$ Stations \\ $288$ Grid cells \end{tabular} &
1949-1999 &
\begin{tabular}{@{}c@{}}Stations: $[0.087, 0.127]$ \\ Gridded: $[0.040, 0.120]$ \end{tabular} \\
\hline

\cite{Papalexiou2013battle} &
\begin{tabular}{@{}c@{}}Annual \\ Maxima of \\ Daily \\ rainfall \end{tabular} &
15137 Stations &
40-163 years &
\begin{tabular}{@{}c@{}}Estimated: $[-0.59, 0.76]$ \\ 90\% ECI: $[-0.11, 0.28]$ \\ Limiting (99\% CI): $[0.0, 0.23]$ \end{tabular} \\
\hline

\cite{Ragulina2017generalized} &
\begin{tabular}{@{}c@{}}Annual \\ Maxima of \\ Daily \\ precipitation\end{tabular} &
1495 Stations &
\begin{tabular}{@{}c@{}}$\geq$99 years \\ with $<5\%$ \\ missing values\end{tabular} &
\begin{tabular}{@{}c@{}}Global mean $\xi$ \\ (95\% CrI) \\ $[0.127,\,0.150]$\end{tabular} \\
\hline

\cite{Roux2023projection} &
\begin{tabular}{@{}c@{}}Annual \\ Maxima of \\ Daily \\ snowfall\end{tabular} &
\begin{tabular}{@{}c@{}}23 massifs \\ French Alps \\ (900-3600 m) \end{tabular} &
\begin{tabular}{@{}c@{}}S2M reanalysis \\ 1959-2019 \end{tabular} &
$[-0.10, 0.17]$ \\
\hline

\end{tabular*}

}
\end{table}

With the shape parameter ranges in mind, we now turn our attention to formalizing strong prior distributions that appropriately span these ranges. \cite{Martins2000generalized} introduces Generalized Maximum Likelihood (GML), a semi-Bayesian framework to estimate the GEV parameters for the hydrologic extreme events. Based on the domain hydrological expertise, they believed that $\xi$ should typically fall within a heavy-tailed range $-0.35 \leq \xi \leq 0$. This range is interpreted as heavy-tailed in \cite{Martins2000generalized} due to their use of a reparameterized form (i.e., a different sign convention) of Equation~\ref{eq:gevpdf}. So, they introduce $Beta(6,9),\quad \xi \in [-0.5,0.5]$ as the prior distribution for the $\xi$ parameter to restrict the unrealistic estimates for the shape parameter. \cite{Yoon2010afull} extend this approach to a complete Bayesian framework with Metropolis-Hastings algorithm called BAYBETA and by changing the prior hyper-parameters as  $Beta(6.4990,8.7927),$ $ \xi \in [-0.5,0.5]$. The section~\ref{section:methodology} presents the framework for incorporating a Beta prior to estimate extreme snow load distributions across multiple geographic locations.

\section{Methodology}\label{section:methodology}

Based on the reviewed literature concerning prior distributions for the GEV shape parameter $\xi$, we adopt the range $\xi \in [-0.1, 0.23]$, which is consistent with findings from precipitation, rainfall, and snowfall studies. This range is motivated by \cite{Roux2023projection}, which shows that $\xi$ can be negative at high elevations for snow extremes, and \cite{Papalexiou2013battle}, which narrows the range of $\xi$ to between 0 and 0.23 for extreme rainfall events. This is a reasonable range because it allows for a broad range of estimated shape parameters, while avoiding the impractical extrapolations that occur in a structural reliability analysis when |$\xi$| > 0.25.

\begin{figure}[htbp]
    \centering
    \begin{subfigure}[b]{0.32\textwidth}
        \centering
            \includegraphics[width=\linewidth]{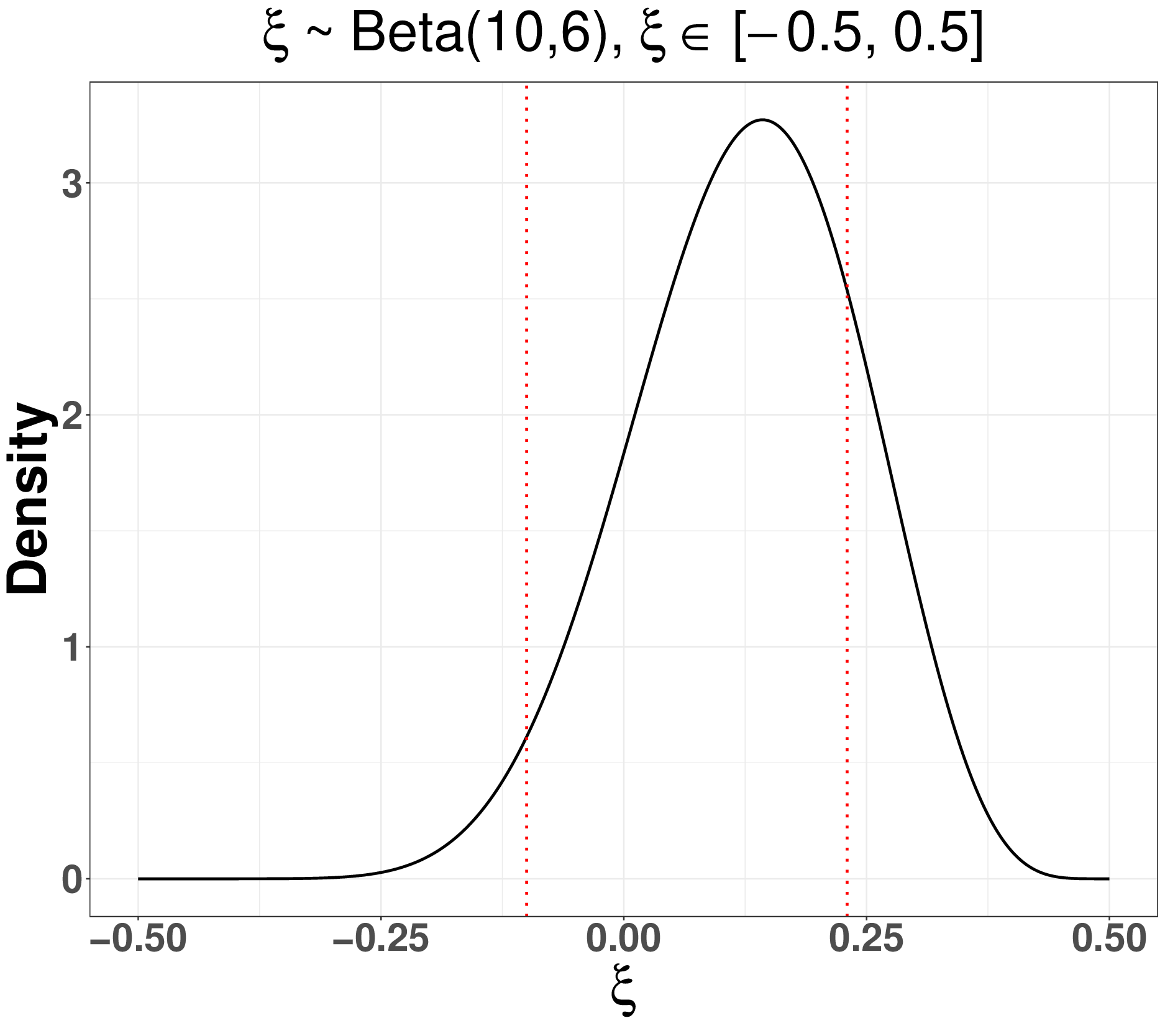}
        \caption{}
    \end{subfigure}
    \hfill
    \begin{subfigure}[b]{0.32\textwidth}
        \centering
        \includegraphics[width=\linewidth]{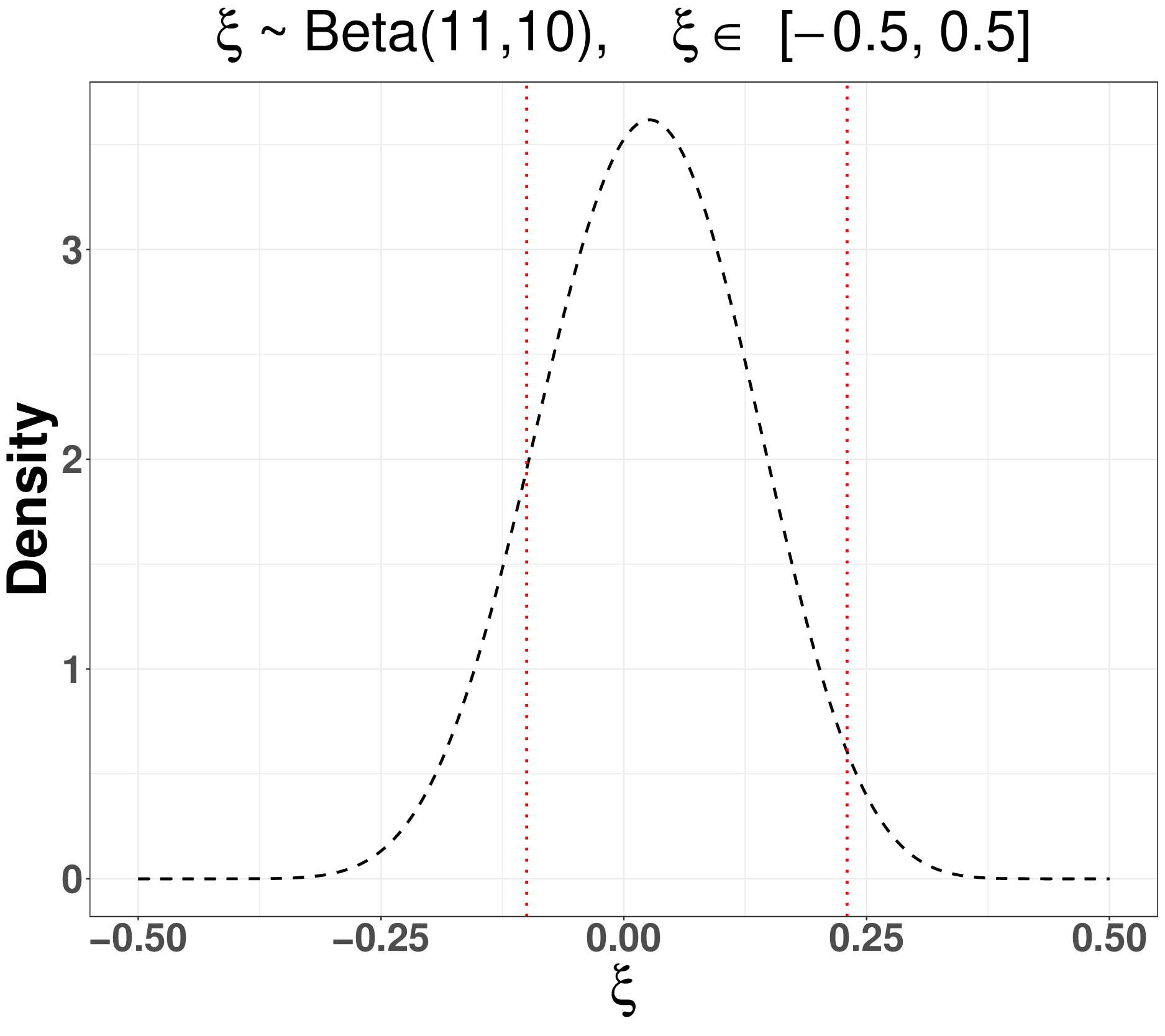}
        \caption{}
    \end{subfigure}
    \hfill
    \begin{subfigure}[b]{0.32\textwidth}
        \centering
        \includegraphics[width=\linewidth]{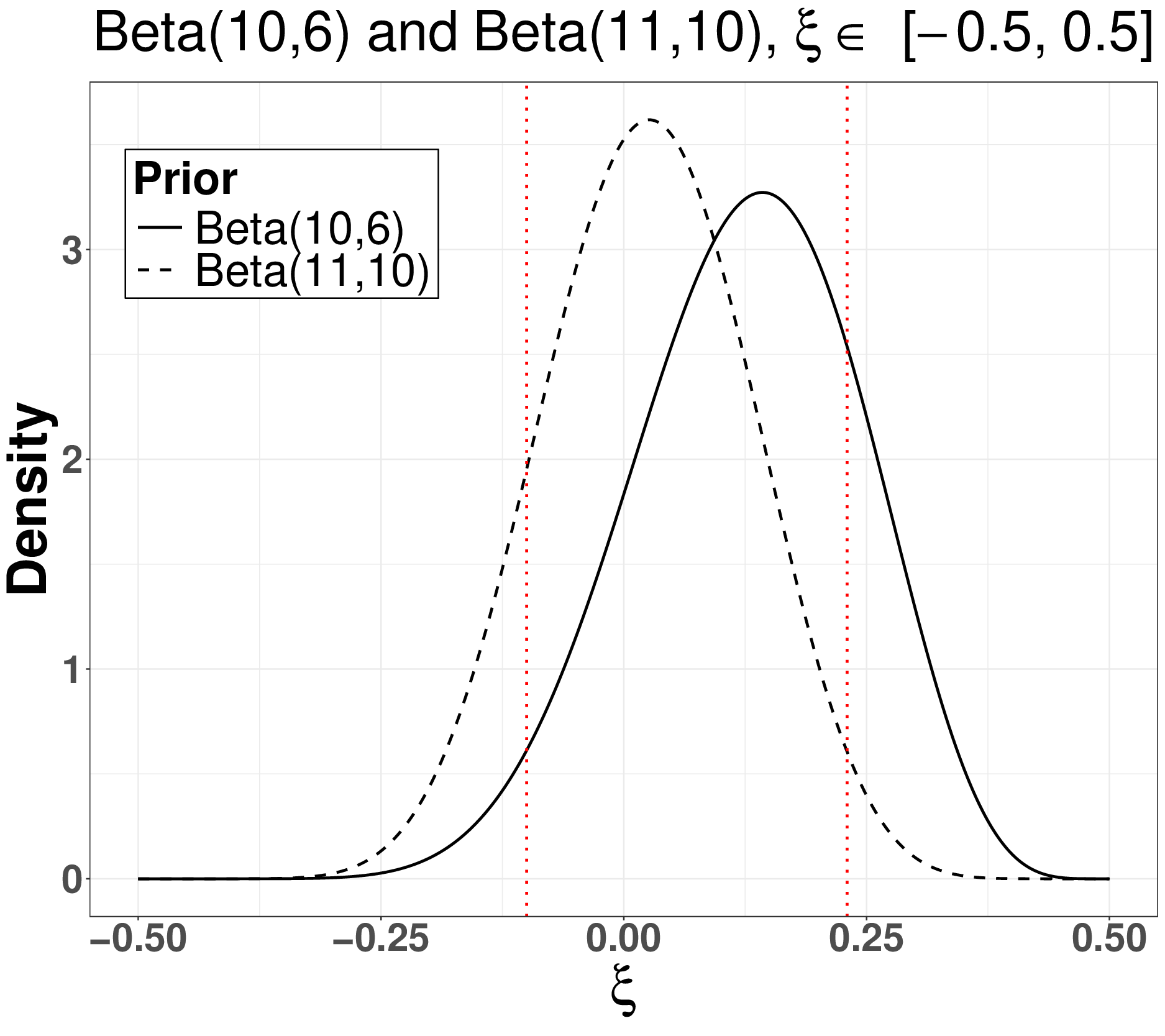}
        \caption{}
    \end{subfigure}
    \caption{(a) Density curve for the baseline prior $Beta(10,6)$; (b) density curve obtained after adjusting the prior to accommodate larger median values, $Beta(11,10)$; (c) comparison of the two priors, illustrating the resulting shift in density. In all three panels, the vertical dotted lines denote the reference values $\xi = -0.1$ and $\xi = 0.23$.}
    \label{fig:betaprior}
\end{figure}

To ensure that the shape parameter prior is constrained to the scientifically justified range, we adopt a Shifted Beta distribution, whose flexible support on a finite interval allows us to encode this prior knowledge directly. Since the defined prior range contains both positive and negative regions, we consider two Beta prior distributions. A positively skewed prior, $Beta(10,6)$ (see Figure~\ref{fig:betaprior}(a)), concentrates most of its density on positive values of the shape parameter and is appropriate for locations with moderate median annual maximum snow loads. On other other hand, we defined our second prior as $Beta(11,10)$ (see Figure~\ref{fig:betaprior}(b)), which is a more symmetric (slightly positively skewed), allowed greater probability mass in the negative region compared to the $Beta(10,6)$ prior. This prior is better suited for locations that experience season long accumulations of snow, which results in larger median annual maximum snow loads as commonly observed in locations with low wintertime temperatures. 

One practical drawback of this globally motivated approach is its reliance on a discrete prior switch. Even small changes in the median annual maximum snow load can abruptly shift the prior from $Beta(10,6)$ to $Beta(11,10)$, leading to unstable inferences around the cutoff point. To address this limitation, we developed a new procedure that smoothly transitions between parameters of the two Beta priors by applying a hyperbolic tangent transformation to the natural logarithm of the median annual maximum snow load. The hyperbolic tangent function, shown in Equation~\ref{eq:tanh}, smoothly maps real numbers to the interval $(-1, 1)$, enabling a continuous and stable interpolation between the two prior specifications.

\begin{equation}\label{eq:tanh}
\text{tanh}(x) = \frac{e^{2x} - 1}{e^{2x} + 1}, \quad x \in \mathbb{R}   
\end{equation}

Since we need to restrict the first prior parameter, denoted by $\eta$, to vary between 10 and 11, and the second prior parameter, denoted by $\kappa$, to vary between 6 and 10, we define smooth transition functions for both parameters. Let $\Tilde{m}$ denote the median annual maximum snow load. The transition functions are then formulated as follows:

\begin{equation}\label{eq:betapriorparm}
\begin{aligned}
\eta &= 0.5 \ \text{tanh}(\log(\Tilde{m})) + 10.5  \\ 
\kappa &= 2 \ \text{tanh}(\log(\Tilde{m})) + 8
\end{aligned}     
\end{equation}

\begin{figure}[htbp]
    \centering
    \begin{subfigure}[b]{0.32\textwidth}
        \centering
            \includegraphics[width=\linewidth]{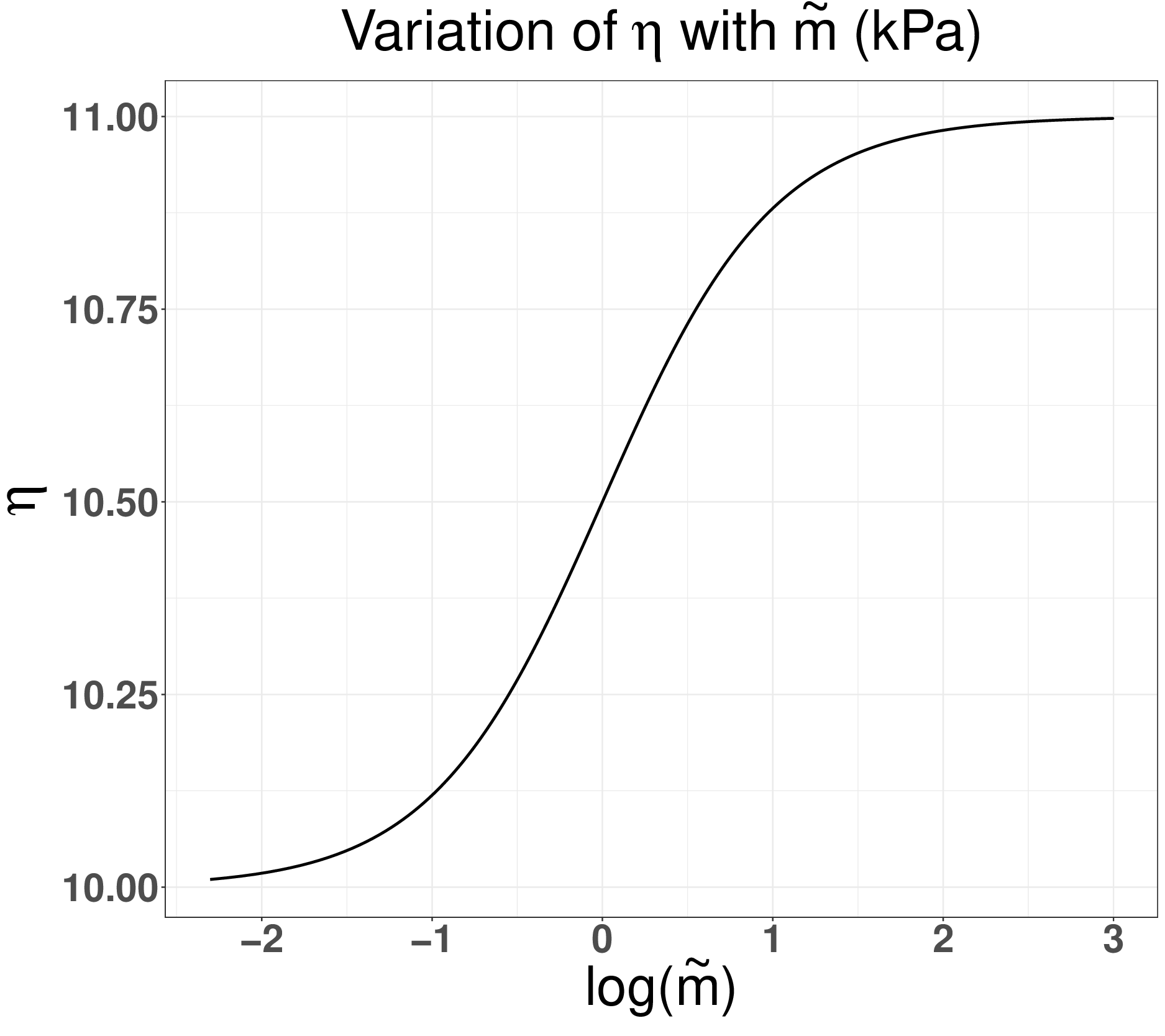}
        \caption{}
    \end{subfigure}
    \hfill
    \begin{subfigure}[b]{0.32\textwidth}
        \centering
        \includegraphics[width=\linewidth]{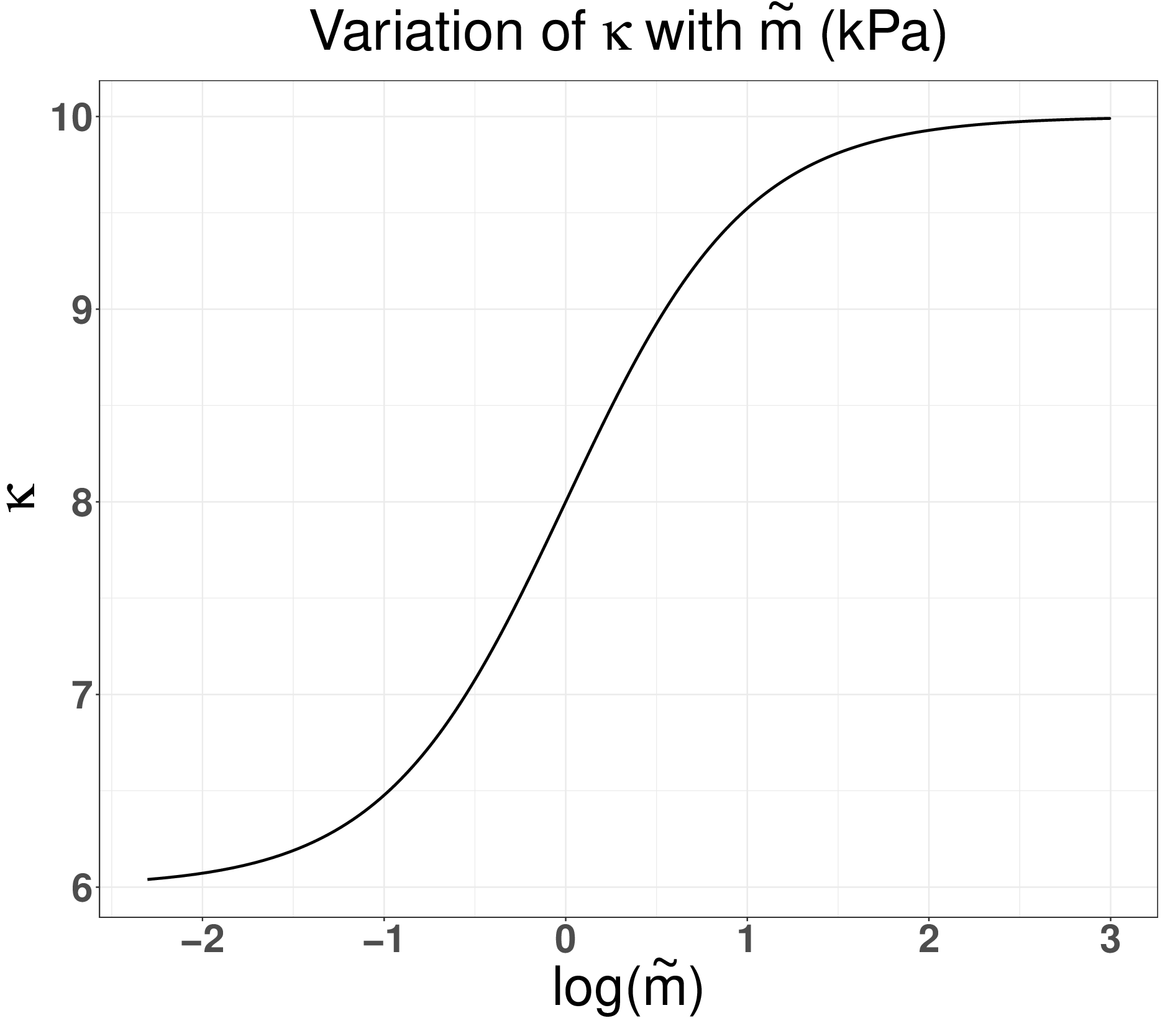}
        \caption{}
    \end{subfigure}
    \hfill
    \begin{subfigure}[b]{0.32\textwidth}
        \centering
        \includegraphics[width=\linewidth]{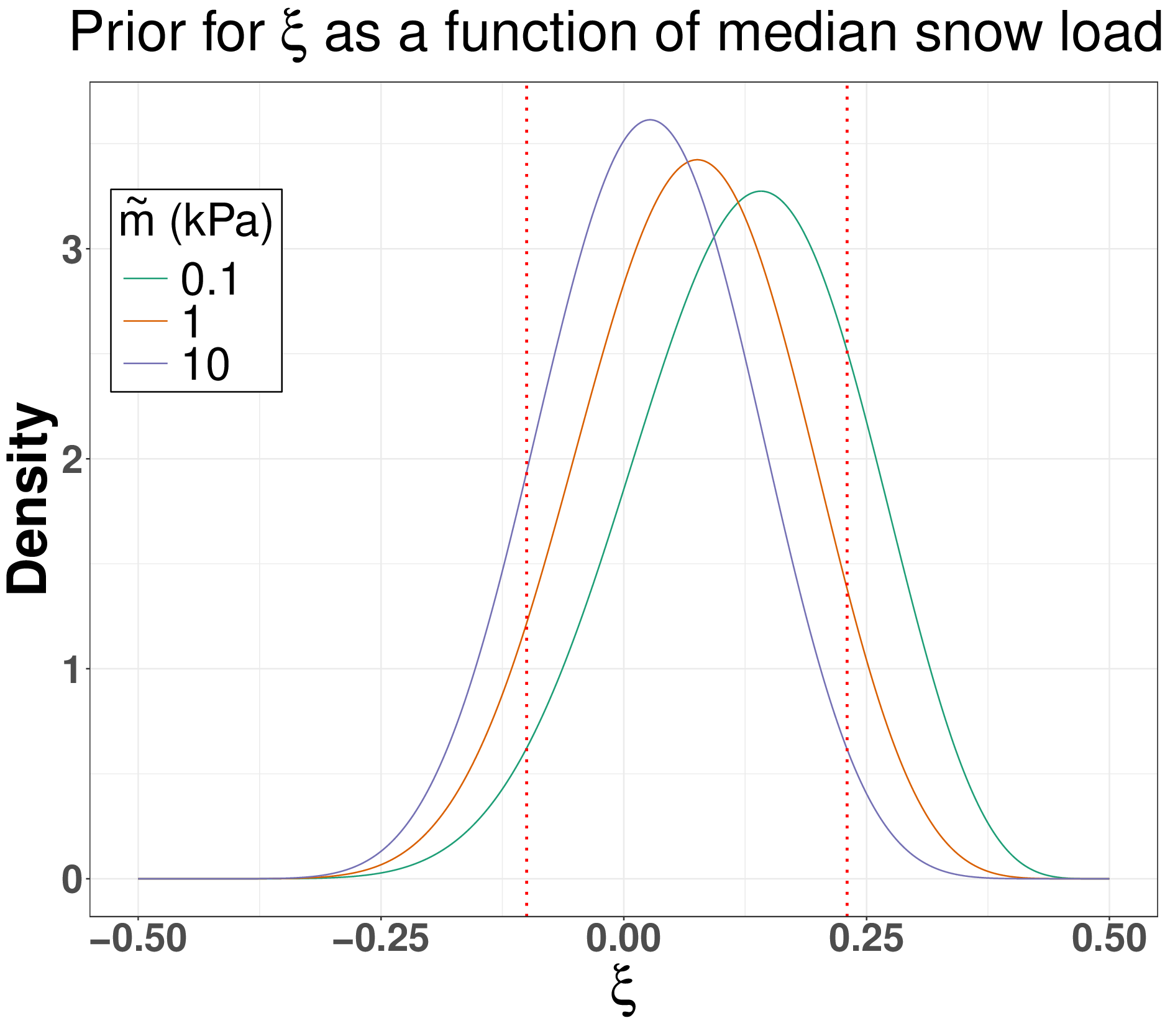}
        \caption{}
    \end{subfigure}
    \caption{(a) Variation of $\eta$ as a function of $\log(\tilde{m})$; (b) Variation of $\kappa$ as a function of $\log(\tilde{m})$; (c) Prior distributions for the $\xi$ corresponding to selected values of the $\tilde{m}$ (kPa). The prior, defined as $\xi \sim Beta(\eta, \kappa), \quad \xi \in [-0.5,0.5]$, transitions smoothly from a positively skewed distribution toward a more symmetric form with increasing snow load severity.}
    \label{fig:sbtanh}
\end{figure}

Based on equation~\ref{eq:betapriorparm}, we define a prior distribution for shape parameter as~$\xi \sim Beta(\eta, \kappa), \quad \xi \in [-0.5,0.5]$. Figure~\ref{fig:sbtanh}(a) and Figure~\ref{fig:sbtanh}(b) illustrate how $\eta$ and $\kappa$ vary as $\tilde{m}$ changes, respectively. Figure~\ref{fig:sbtanh}(c) shows how the prior distribution of the $\xi$, varies with $\tilde{m}$, expressed in kilopascals. For locations with relatively small $\tilde{m}$, the prior is positively skewed and as $\tilde{m}$ increases, the tanh(.) transformation induces a smooth transition rather than abruptly toward a more symmetric prior, allowing increasing mass in the negative region of the shape parameter. Additionally, this method avoids a fixed threshold on $\tilde{m}$ and ensures stable inference across a wide range of climatological conditions.

\subsection{Bayesian Formulation}

\subsubsection{HMC Formulation for Posterior Inference in the GEV Model}

In this study, we use the HMC algorithm to estimate the parameters of the GEV distribution. Since the scale parameter $\sigma$ is defined only on the positive real line, we re-parameterize it using $\delta = \log(\sigma)$ to generate posterior samples. This transformation maps $\sigma$ to the entire real line, enabling more efficient and stable sampling within the HMC framework (see Equation~\ref{eq:gevpdf}). Since the GEV distribution exhibits complex geometric properties, we sample the auxiliary momentum variables $\boldsymbol{p}$ from $\mathcal{N}_d(\boldsymbol{0}, \boldsymbol{I})$, where $\boldsymbol{I}$ is the identity matrix. We can now define the Hamiltonian function for the GEV model as follows:

\begin{equation}\label{eq:hmcfgev}
\begin{aligned}
H(\mu,\delta,\xi, \boldsymbol{p};\boldsymbol{x}) = -\log(P(\mu,\delta,\xi|\boldsymbol{x})) + \frac{\boldsymbol{p}' \boldsymbol{p}}{2}  
\end{aligned}    
\end{equation}

Where
\begin{equation}\label{eq:loglik}
\begin{aligned}
\log(P(\mu,\delta,\xi|\boldsymbol{x})) \propto \ell(\mu,\delta,\xi;\boldsymbol{x}) + \log(f_{M}(\mu)) + \log(f_{\Delta}(\delta)) + \log(f_{\Xi}(\xi))
\end{aligned}    
\end{equation}

\begin{itemize}
    \item $\ell(\mu,\delta,\xi;\boldsymbol{x}) :=$ log likelihood of GEV density function.
    \item $f_{M}(\mu) :=$ Prior density function for $\mu$.
    \item $f_{\Delta}(\delta) :=$ Prior density function for $\delta$.
    \item $f_{\Xi}(\xi) :=$ Prior density function for $\xi$.
\end{itemize}

Now we can write the gradient for leapfrog integration in HMC as,

\begin{equation}\label{eq:gradhmc}
\begin{aligned}
\frac{\partial H(\mu,\delta,\xi, \boldsymbol{p})}{\partial \boldsymbol{p}} &= \boldsymbol{p} \\
-\frac{\partial H(\mu,\delta,\xi, \boldsymbol{p})}{\partial [\mu,\delta,\xi]} &= \nabla_{[\mu,\delta,\xi]} \log(P(\mu,\delta,\xi|\boldsymbol{x}))
\end{aligned}    
\end{equation}

We implemented this version of the Hamiltonian Monte Carlo (HMC) algorithm with dual averaging, tailored to the GEV distribution and using analytical gradients, as a custom \texttt{R} package called \texttt{extremMHMC}, a copy of which is included in the manuscript's supplementary material. The all core computational routines are written in \texttt{Rcpp} \citep{Eddelbuettel2011rcpp} to improve computational efficiency. This package was developed using R 4.5.0 \citep{Rlang}. 

\subsubsection{Priors for GEV parameters}

The shape parameter $\xi$ requires a constrained prior, a Beta prior is specified for $\xi$ with support $[-0.5, 0.5]$ that reflects physically meaningful tail behavior for annual maximum snow loads. In contrast, both the location parameter $\mu$ and the transformed scale parameter $\delta = \log(\sigma)$ are defined on the entire real line. Therefore, independent Normal priors are assigned to $\mu$ and $\delta$, providing a flexible and computationally convenient specification for parameters without natural bounds. These prior selections ensure that the resulting posterior distribution is well-behaved, adheres to model constraints, and remains compatible with efficient gradient-based sampling in the HMC framework.

\begin{equation}\label{eq:priors}
\begin{aligned}
\mu &\sim \mathcal{N}(\Tilde{m}, 100^2) \\
\delta &\sim \mathcal{N}(0, 100^2) \\
\xi &\sim Beta(\eta, \kappa), \quad \xi \in [-0.5,0.5]
\end{aligned}    
\end{equation}

Occasionally, $\xi$ falls outside the admissible interval $[-0.1, 0.23]$ when estimating parameters using the priors in equation \ref{eq:priors}. To resolve this, we define a new prior for $\xi$ to effectively bound the estimated shape parameters within this range and re-estimate the parameters.

\begin{equation}\label{eq:priorsnew}
\begin{aligned}
\mu &\sim \mathcal{N}(\Tilde{m}, 100^2) \\
\delta &\sim \mathcal{N}(0, 100^2) \\
\xi^* &\sim \begin{cases}
\mathcal{N}(-0.095, 0.01^2) , &\xi <  -0.1 \\
\\
\mathcal{N}(0.225, 0.01^2) , &\xi > 0.23
\end{cases} 
\end{aligned}    
\end{equation}

\subsubsection{Validate Posterior parameter chain}

A single HMC run produces posterior chains for all three parameters $\mu$, $\delta$ and $\xi$ simultaneously, so it is necessary to assess the convergence of each chain individually. To assess convergence, we rely on two widely used statistical diagnostics in Bayesian statistics, the Potential Scale Reduction Factor ($\hat{R}$) and the Effective Sample Size (ESS).

Potential Scale Reduction Factor also known as Gelman-Rubin diagnostic introduced by \cite{Gelman1992inference} to monitor the convergence of Markov Chain Monte Carlo (MCMC) simulations. To asses the convergence using $\hat{R}$ we need multiple chains (at least three chains) per parameter with same number of posterior parameter samples. The $\hat{R}$ statistic is computed by comparing the within-chain variance to the between-chain variance. Values of $\hat{R}$ close to 1.0 indicate that the chains have mixed well and converged to the target distribution, whereas values substantially greater than 1.0 suggest that additional sampling is required.

While the ESS metric assesses the quality of a single MCMC chain, the concept of ESS lacks a unique point of origin and, its modern use in MCMC became widely established following the work of \cite{Geyer1992practical}. The ESS metric quantifies how many independent samples from the target distribution would contain the same amount information as the correlated samples from the single Markov chain. A low ESS suggests poor sampling efficiency, implying that the MCMC chain yields less reliable parameter estimates due to strong autocorrelation and insufficiently independent samples. 

In our study, to assess the convergence, we are using a different variation of classical $\hat{R}$ and ESS introduced by \cite{Vehtari2021rank} for MCMC chain validation. Because classical $\hat{R}$ and ESS are based on variance and autocorrelation estimates that assume finite mean and variance of the target posterior distribution, but this assumption fails for heavy-tailed distributions or distributions with infinite variance (e.g., Cauchy). To overcome this issue, their study introduced "Rank-Normalization" combining all chains. Also, \cite{Vehtari2021rank} divided the ESS into two parts as Bulk-ESS and Tail-ESS. According to their study, if improved rank-normalized $\hat{R} \leq 1.05$ and, both the combined-chain Bulk ESS and combined-chain Tail ESS are greater than 400, then the chains have converged to the same invariant distribution and are well mixed, providing reliable estimates. The computation of this version of $\hat{R}$, as well as the bulk-ESS and tail-ESS, is implemented in the \texttt{rstan} \citep{rstan} library in \texttt{R}, and we used this library to obtain all convergence diagnostics reported in this work. For more details regarding the calculation of $\hat{R}$ and ESS, please refer to Appendix~\ref{appendix:A2}.

\subsubsection{Summary of the Parameter Estimation Procedure}

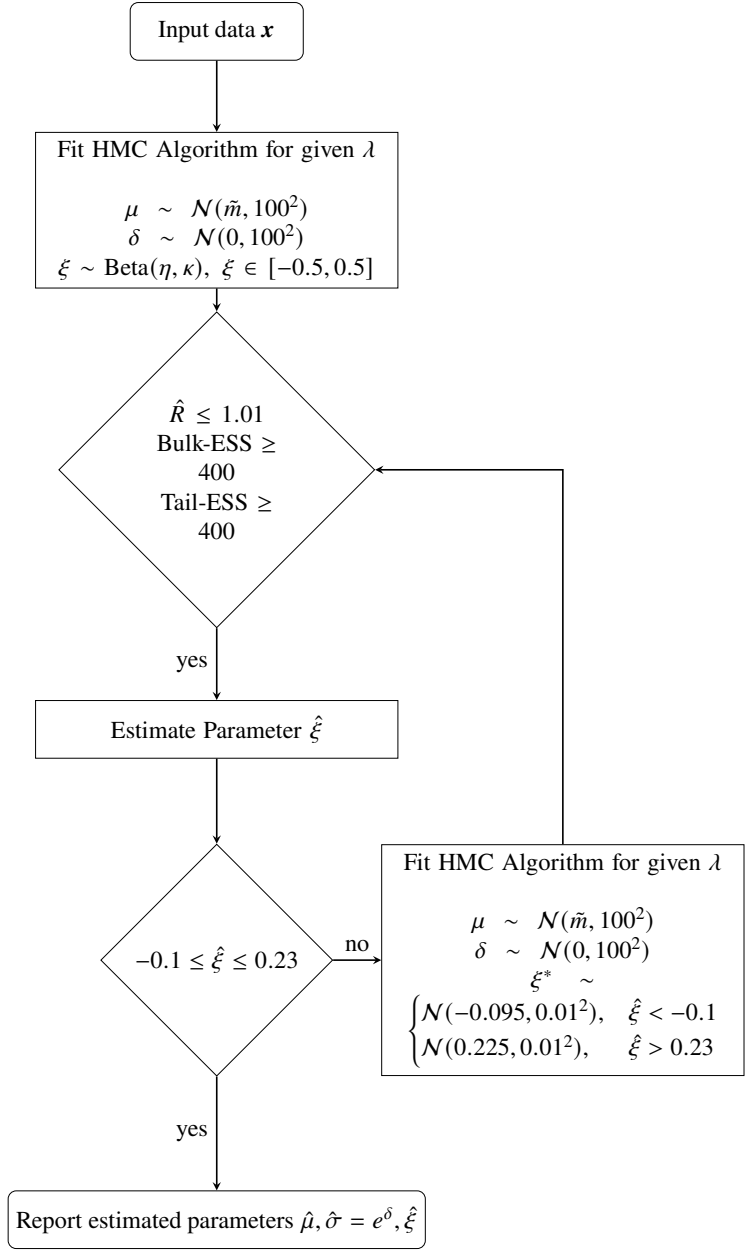
\begin{figure}[htbp]
\centering
\resizebox{0.6\textwidth}{!}{
\begin{tikzpicture}[node distance=3cm]

\node (start) [startstop] {Input data $\boldsymbol{x}$};

\node (pro1) [process1, below of = start, yshift=-0.1cm] {
Fit HMC Algorithm for given $\lambda$

\

\(\mu \sim \mathcal{N}(\tilde{m},100^2)\)

\(\delta \sim \mathcal{N}(0,100^2)\)

\(\xi \sim \text{Beta}(\eta,\kappa),\ \xi \in [-0.5,0.5]\)
};

\node (dec1) [decision1, below of=pro1, yshift=-1.5cm] {\(\hat{R} \leq 1.01 \)

\(\text{Bulk-ESS} \geq 400 \)

\(\text{Tail-ESS} \geq 400 \)
};

\node (pro2a) [process1, below of=dec1, yshift=-1.5cm] {Estimate Parameter $\hat{\xi}$};

\node (dec2) [decision2, below of=pro2a, yshift=-1.0cm] {\( -0.1 \leq \hat{\xi} \leq 0.23 \)};

\node (out) [startstop, below of=dec2, yshift=-1.5cm] {Report estimated parameters

\( \hat{\mu}, \hat{\sigma} = e^\delta, \hat{\xi} \)
};

\node (pro2) [process1, right of = dec2, xshift= 3cm] {
Fit HMC Algorithm for given $\lambda$

\

\(\mu \sim \mathcal{N}(\tilde{m},100^2)\)

\(\delta \sim \mathcal{N}(0,100^2)\)

\(\xi^* \sim \begin{cases}
\mathcal{N}(-0.095, 0.01^2) , &\hat{\xi} <  -0.1 \\
\mathcal{N}(0.225, 0.01^2) , &\hat{\xi} > 0.23
\end{cases}\)
};

\draw [arrow] (start) -- (pro1);
\draw [arrow] (pro1) -- (dec1);
\draw [arrow] (dec1) -- node[anchor=east] {yes} (pro2a);
\draw [arrow] (pro2a) -- (dec2);
\draw [arrow] (dec2) -- node[anchor=east] {yes}(out);
\draw [arrow] (dec2) -- node[anchor=south] {no}(pro2);
\draw [arrow] (pro2) |- (dec1);

\end{tikzpicture}
}
\caption{Flowchart of the parameter estimation methodology.}
\label{fig:flowchart}
\end{figure}

Estimating the parameters under the Squared Error Loss (SEL) corresponds to computing the posterior mean of each parameter. For the estimation scale parameter $\sigma$, the posterior samples of $\delta$ are first transformed using $\sigma = e^{\delta}$ before obtaining the posterior estimate. Before applying the Normal prior for $\xi$ (see equation~\ref{eq:priorsnew}), several trajectory lengths ($\lambda$) are tested to ensure that, for some choice of $\lambda$, the estimated values of $\xi$ with a shifted Beta prior remain within the desired range $[-0.1,0.23]$. In practice, $\lambda$ is typically varied over the interval $[0,3]$ but, sometimes $\lambda$ can be as high as 10. Figure \ref{fig:flowchart} flowchart shows the idea of estimating the parameters.

\subsection{Estimate Reliability Targeted Snow Loads (RTSL)}

After estimating the GEV parameters, the next step is to compute Reliability Targeted Snow Loads (RTSLs). Estimating RTSLs requires modeling the relationship between the hazard (e.g., snow), the resistance members (e.g., a steel beam), and the design code. RTSLs are site-calibrated snow loads, determined so that the probability of failure meets a prescribed Reliability index ($\beta$). The target probabilities of failure, as prescribed in Chapter 1 of the ASCE 7 Standard \citep{ASCE722}, are very small. For example, a typical "Risk Category II" structure would be associated with $\beta=3.0$ which corresponds to a 0.13\% chance of building failure due to excess loading in a 50 year period. Such small probabilities of failure requires severe extrapolations into the tail of the GEV distribution describing the annual maximum snow loads. As such, the calculated RTSLs are largely governed by the estimates of the GEV shape parameter.

The RTSL estimation process is based on a probability-based method for calculating load factors that was introduced by \cite{Ellingwood1980development}. Various researchers recalibrate this framework to understand the distribution of load factors and the distribution of resistance members.  \cite{DeBock2016colorado} introduced a method to calculate the site-specific RTSLs based on data from over 300 snow measurement stations across Colorado. Their study showed that, in order to maintain a uniform level of safety, snow loads would need to fall slightly in the higher elevation regions of Colorado and increase, sometimes substantially on the Eastern Plains of Colorado. The reason for this is because the 50-year MRI snow load, which used to be the basis for design, does not well approximate RTSLs in a region like Colorado where the annual maximum snow load generating process is so variable in the mountains (with accumulations over months) vs the plains (with accumulations over just a few days). Later, \cite{Bean2021SnowLoad} expanded this approach by computing site-specific reliability-targeted loads (RTSLs) across the United States.

Estimation of RTSLs, as emulated in this paper, are based on the multi-step process described further in \cite{Bean2021SnowLoad} and summarized below for convenience: 
\begin{enumerate}
    \item Define a target design scenario and set up the resulting design equation using information from the ASCE 7 standard.
    \item Simulate the site-specific 50-year ground snow loads using the GEV distribution, along with a ground to roof conversion factor that is conditional upon the ground snow load.
    \item Simulate the dead load and structural resistance for the target design scenario.  
    \item Count the number of simulated failures, and increase the structural resistance until the simulated failures in the simulation until the simulated failures fall below a target amount. 
\end{enumerate}

The key highlight from this process that is relevant to this paper is the central role that the ground snow load GEV distribution plays in calculating RTSLs. As shown in Section \ref{section:simulation_study}, even small changes in the estimated shape parameter can result in proportionally large changes in the estimated RTSLs.

\section{Simulation Study}\label{section:simulation_study}

\subsection{Impact of GEV Parameters on estimating RTSL}

We conduct a simulation study to examine the impact of the shape parameter on RTSLs relative to the other two parameters. Four weather stations located in different regions of the United States are selected: Washington, DC; Denver, CO; Duluth, MN; and Sierra City, CA. For each location, RTSLs are first estimated using smooth L-moments \citep{Bean2021SnowLoad}. To obtain the smooth L-moments, first standard L-moments were used to estimate the parameters, and then Generalized Additive Models (GAM) were used to spatially smooth the shape parameter as a function of the median of annual maxima. During the estimation phase, the shape parameter was limited to fall between 0 and 0.25 for both the hindcast and future scenario periods to avoid unrealistic estimates of RTSLs.

In this simulation study, Location and Scale parameters are then independently increased and decreased by 10\%, 25\%, and 50\%. For the Shape parameter instead of percentage changes, we used a additive effects $\pm 0.01$, $\pm 0.025$ and $\pm 0.05$ because $\xi$ is generally small in magnitude and even modest shifts can strongly affect the high RTSLs. Additionally, we changed one parameter, while holding the remaining two parameters fixed at their original smooth L-moment estimates, and the RTSL is re-estimated for each case (see Table~\ref{tab:rtlsmest}). For each parameter combination, the RTSL is estimated five times, and the median of these estimates is used to reduce the influence of small fluctuations in the estimates RTSLs due to the simulation process. The relative change in the RTSL is then computed with respect to the original estimate for each parameter combination.

\begin{table}[h]
\centering
\caption{Estimated parameters and Reliability Targeted Snow Load (RTSL) values for selected locations using smooth L-moments.}
\label{tab:rtlsmest}

\begin{tabular}{c c c c c}
\hline
Location & $\mu$ & $\sigma$ & $\xi$ & RTSL (kPa) \\
\hline
Denver, CO & 0.20 & 0.09 & 0.25 & 2.55 \\
\hline
Washington, DC & 0.20 & 0.13 & 0.25 & 3.46 \\
\hline
Duluth, MN & 0.97 & 0.43 & 0.00 & 4.36 \\
\hline
Sierra City, CA & 2.18 & 1.42 & 0.04 & 15.56 \\
\hline
\end{tabular}
\end{table}

\begin{table}[h]
\centering
\caption{Table showing how RTSL estimates change relative to their original value when a one parameter changes and other two remain constant. RTSL represent the Reliability Targeted Snow Loads with unit kPa, and RRC represent the percentage of Relative RTSL Change. }
\label{tab:relchange}

{\scriptsize

\begin{tabular*}{\textwidth}{@{\extracolsep{\fill}} c c c c c c c c c c @{}}
\hline
\multirow{2}{*}{Parameter} &
\multirow{2}{*}{\begin{tabular}{@{}c@{}}Relative \\ Change\end{tabular}} &
\multicolumn{2}{c}{Denver} &
\multicolumn{2}{c}{Washington} &
\multicolumn{2}{c}{Duluth} &
\multicolumn{2}{c}{Sierra City} \\
\cline{3-4}\cline{5-6}\cline{7-8}\cline{9-10}
& & RTSL (kPa) & RRC (\%) & RTSL (kPa) & RRC (\%) & RTSL (kPa) & RRC (\%)  & RTSL (kPa) & RRC (\%) \\
\hline

\multirow{6}{*}{$\mu$}
 & -50\% & 2.44 & -4.3 & 3.41 & -1.4 & 3.81 & -12.6 & 14.55 & -6.5 \\
 & -25\% & 2.47 & -3.1 & 3.43 & -0.9 & 4.08 & -6.4 & 15.15 & -2.6 \\
 & -10\% & 2.53 & -0.8 & 3.49 & 0.9 & 4.26 & -2.3 & 15.58 & 0.1 \\
 &  10\% & 2.54 & -0.4 & 3.50 &  1.2 & 4.48 & 2.8 & 15.98 & 2.7 \\
 &  25\% & 2.57 &  0.8 & 3.53 &  2.0 & 4.67 & 7.1 & 16.42 & 5.5 \\
 &  50\% & 2.62 &  2.7 & 3.56 &  2.9 & 4.95 & 13.5 & 16.96 & 9.0 \\
\hline

\multirow{6}{*}{$\sigma$}
 & -50\% & 1.37 & -46.3 & 1.79 & -48.3 & 2.70 & -38.1 & 9.00 & -42.2 \\
 & -25\% & 1.94 & -23.9 & 2.61 & -24.3 & 3.52 & -19.3 & 12.40 & -20.3 \\
 & -10\% & 2.30 & -9.8 & 3.16 & -8.7 & 4.03 & -7.6 & 14.42 & -7.3 \\
 &  10\% & 2.78 & 9.0 & 3.84 & 11.0 & 4.71 & 8.0 & 17.18 & 10.4 \\
 &  25\% & 3.19 & 25.1 & 4.42 & 27.7 & 5.23 & 20.0 & 19.20 & 23.4 \\
 &  50\% & 3.81 & 49.4 & 5.27 & 52.3 & 6.08 & 39.4 & 22.58 & 45.1 \\
\hline

\multirow{6}{*}{$\xi$}
 & -0.050 & 1.94 & -23.9 & 2.60 & -24.9 & 3.85 & -11.7 & 13.44 & -13.6 \\
 & -0.025 & 2.19 & -14.1 & 3.01 & -13.0 & 4.08 & -6.4 & 14.48 & -6.9 \\
 & -0.010 & 2.40 & -5.9 & 3.31 & -4.3 & 4.26 & -2.3 & 15.27 & -1.9 \\
 & +0.010 & 2.69 & 5.5 & 3.73 & 7.8 & 4.48 & 2.8 & 16.32 & 4.9 \\
 & +0.025 & 2.94 & 15.3 & 4.07 & 17.6 & 4.70 & 7.8 & 17.23 & 10.7 \\
 & +0.050 & 3.49 & 36.9 & 4.82 & 39.3 & 5.09 & 16.7 & 19.02 & 22.2 \\
\hline

\end{tabular*}
}
\end{table}

According to the Table~\ref{tab:relchange}, for all four locations, percentage changes in the $\mu$ parameter result in relative small changes to the estimated RTSL. The scale parameter's effect on RTSL values is much more substantial than the location parameter. Except for Duluth, in all three other locations, relative percentage increase or decrease in RTSL is very similar to the relative percentage changes in $\sigma$. For example, in Denver, when the relative change in $\sigma$ is 50\%, the RTSL change is 49.4\%. In Duluth, where the shape parameter is equal to zero, the relative change in the RTSLs are slightly muted relative to the other locations.

For the shape parameter, we used additive changes rather than percentage changes (multiplicative changes), because, unlike $\mu$ and $\sigma$, $\xi$ is usually small in magnitude and strongly controls tail heaviness. Changing the $\xi$ by $\pm 0.01$ produces a relatively small percentage increase or decrease in RTSLs for all selected locations. Across all locations, adding 0.05 results in a larger effect in RTSL than reducing the same amount, specifically in Denver (36.9\%) and Washington, DC (39.3\%). This is because adding 0.05 to the original $\xi$ value of those two locations exceeds 0.25. RTSLs in Duluth show less sensitivity to shifts $\xi$, compared to the other three locations for the same magnitude of change. For Sierra City, adding 0.05 increases the RTSL by around 22.2\%, but this is not as large as Denver and Washington. One reason this could be is even adding 0.05, Sierra City is still within the physically plausible range of the shape parameter ($|\xi| < 0.25$).

\begin{figure}[h]
    \centering
    \includegraphics[width=25pc, height = 25pc]{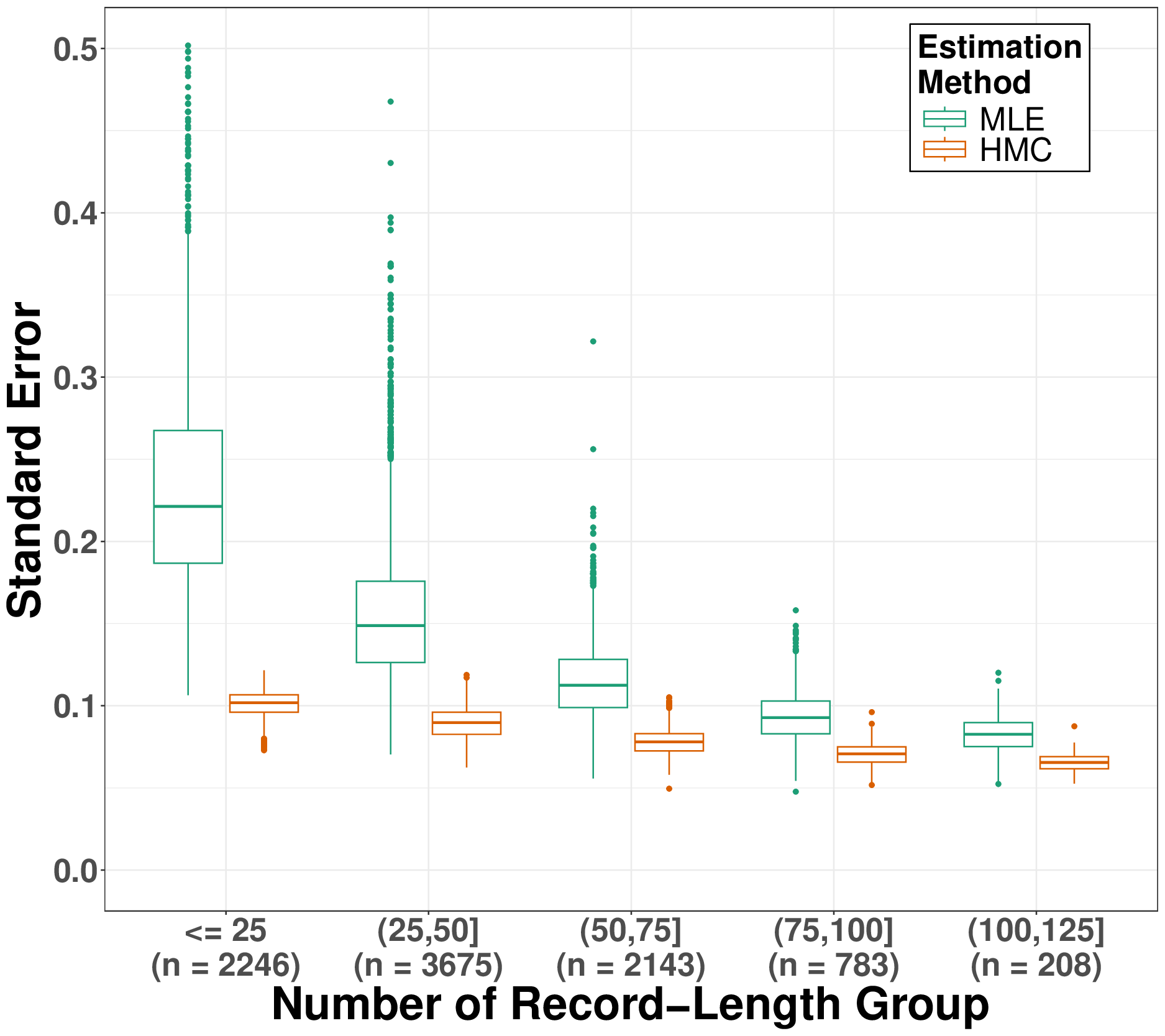}
    \caption{Boxplots showing how the standard error distribution varies across record-length groups when estimating the shape parameter with the MLE approach and the Bayesian HMC approach with the Beta prior. Here, MLE represents the Maximum Likelihood Method, HMC represents the Bayesian HMC approach with the Beta prior, and $n$ represents the number of stations in each record-length group.}
    \label{fig:ase_rl}
\end{figure}

While both the scale and shape parameters largely influence RTSL estimation, the scale parameter is of greatest concern due to the natural variability of the estimate. Additionally, to demonstrate the importance of accurately estimating the shape parameter, we computed the standard errors of all three GEV parameters using the Maximum Likelihood Estimation (MLE) method implemented in the \texttt{extRemes} package \citep{Gilleland2016extRemes} in \texttt{R}. The data used for this MLE estimation are obtained from \cite{Bean2021SnowLoad}. Additional details about the dataset are provided in Section~\ref{section:application}. During estimation, some stations (approximately 6.7\%) produced parameter estimates with $|\xi| \geq 0.5$, indicating unstable or unrealistic fits and these stations were removed from the analysis. Based on the remaining stations, the median standard errors were 0.03, 0.02, and 0.14 for $\mu$, $\sigma$ and $\xi$, respectively. This shows that the median standard error of $\xi$ is approximately 4.7 and 7 times larger than those of $\mu$ and $\sigma$, respectively, indicating considerably greater variability in estimating the shape parameter of the GEV distribution. In Table~\ref{tab:relchange}, we explored perturbations of $\xi$ up to $\pm 0.05$ in magnitude, but the median standard error of $\xi$ is approximately 2.8 times larger than this maximum perturbation (see Figure~\ref{fig:ase_rl} to understand the relationship between the standard error and the sample size). This supports the need for estimation methods that restrict the variance of the $\xi$ parameter. Additionally, Figure~\ref{fig:ase_rl} illustrates that our Bayesian HMC approach with the Beta prior substantially reduces standard errors compared with the naive frequentist MLE, with the largest reductions for stations with the shortest record lengths. The posterior standard deviations from the Bayesian HMC approach are not directly comparable to the standard errors obtained under MLE, even though both serve as analogous measures of parameter uncertainty. However, we want to emphasize that using an informative prior substantially reduces posterior uncertainty in the GEV shape parameter, which is well known to be unstable at short record lengths.

\subsection{Comparison of Frequentist and Bayesian Estimation of the GEV Shape Parameter}

\begin{figure}[h]
    \centering
    \includegraphics[width=30pc, height=30pc]{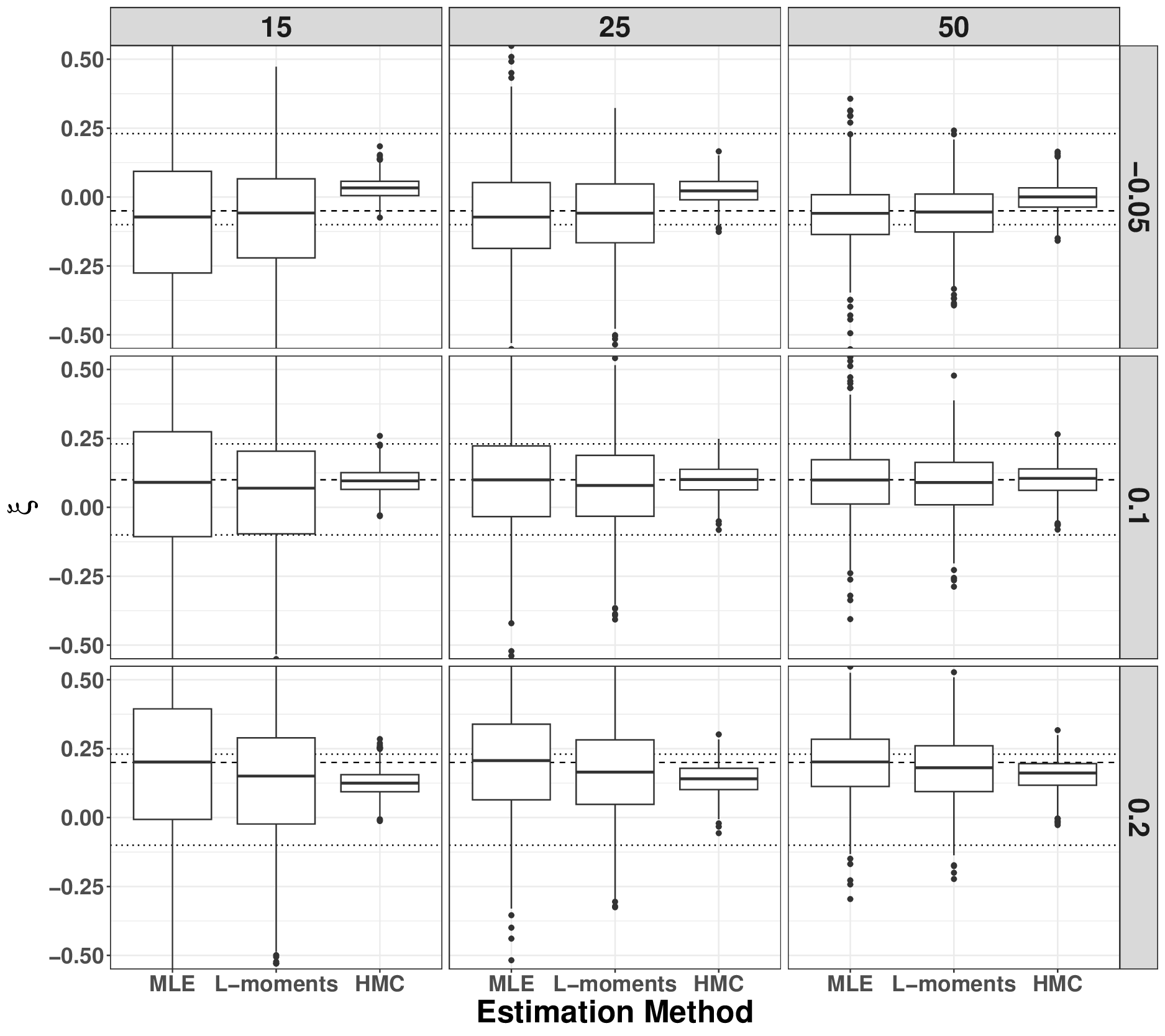}
    \caption{Boxplots of GEV shape parameter estimates from HMC with a $Beta$ prior, L-moments, and MLE across sample sizes (columns) and true shape parameters (rows). Each panel is based on 1000 replications. The dotted horizontal lines indicate the desired range $[-0.1$,$0.23]$, and the dashed horizontal line marks the true shape parameter for that row.}
    \label{fig:sim_results}
\end{figure}

\begin{table}[h]
\caption{RMSE of GEV shape parameter estimates for each $\xi$ and sample size $n$ across the three estimation methods  based on 1000 replications.}
\label{tab:RMSE_simstudy}
\begin{tabular*}{\textwidth}{@{\extracolsep{\fill}}llccccc@{}}
\hline
\multirow{2}{*}{$\xi$} & \multirow{2}{*}{$n$} & \multicolumn{3}{c}{RMSE per Estimation Method} & \multicolumn{2}{c}{Relative RMSE Decrease (\%)} \\
\cline{3-5} \cline{6-7} &  & L-moments & MLE & HMC & L-moments vs HMC & MLE vs HMC \\
\hline
\multirow{3}{*}{-0.05} 
 & 15 & 0.22 & 0.33 & 0.09 & 58 & 73 \\
 & 25 & 0.15 & 0.19 & 0.09 & 40 & 53 \\
 & 50 & 0.10 & 0.12 & 0.07 & 30 & 42 \\
\hline
\multirow{3}{*}{0.00}
 & 15 & 0.22 & 0.33 & 0.07 & 68 & 79 \\
 & 25 & 0.16 & 0.19 & 0.07 & 56 & 63 \\
 & 50 & 0.11 & 0.12 & 0.06 & 45 & 50 \\
\hline
\multirow{3}{*}{0.05}
 & 15 & 0.22 & 0.33 & 0.05 & 77 & 85 \\
 & 25 & 0.16 & 0.20 & 0.06 & 62 & 70 \\
 & 50 & 0.11 & 0.12 & 0.06 & 45 & 50 \\
\hline
\multirow{3}{*}{0.10}
 & 15 & 0.23 & 0.33 & 0.04 & 83 & 88 \\
 & 25 & 0.16 & 0.20 & 0.05 & 69 & 75 \\
 & 50 & 0.11 & 0.13 & 0.06 & 45 & 54 \\
\hline
\multirow{3}{*}{0.15}
 & 15 & 0.24 & 0.35 & 0.06 & 75 & 83 \\
 & 25 & 0.17 & 0.20 & 0.06 & 65 & 70 \\
 & 50 & 0.12 & 0.13 & 0.06 & 50 & 54 \\
\hline
\multirow{3}{*}{0.20}
 & 15 & 0.24 & 0.36 & 0.09 & 62 & 75  \\
 & 25 & 0.17 & 0.21 & 0.08 & 53 & 62  \\
 & 50 & 0.13 & 0.14 & 0.07 & 46 & 50  \\
\hline
\end{tabular*}
\end{table}

To evaluate the performance of the HMC algorithm under a $Beta$ prior for the shape parameter $\xi$, we conducted a series of simulation studies across multiple shape values and sample sizes. These results were compared against the corresponding L-moments and maximum likelihood estimators (MLE). Because the prior on $\xi$ is constrained, a some amount of bias in the Bayesian estimates is expected relative to frequentist estimates. This behavior arises because Bayesian estimators generally shrink the posterior estimates toward regions with higher prior mass (the prior mean), particularly when the prior is informative or the data are limited. Another reason for this small amount of bias in Bayesian methods is more generally frequentist approaches often emphasize unbiasedness and variance of estimators but, Bayesian approaches primarily focused on minimizing the posterior expected loss under a specified prior and loss function (e.g. Squared Error Loss, Absolute Error Loss) \citep{Berger1985Statistical}. However, our primary focus is on reducing posterior uncertainty and lowering Root Mean Square Error (RMSE), both of which are essential for obtaining consistent and practically useful estimates of the shape parameter in extreme value analysis.

We generated 1,000 replications of samples with sizes $n = 15, 25, 50$ from a zero-truncated (i.e., simulated values are bounded below by zero) GEV distribution under six shape parameters: $\xi \in \{-0.05, 0.10, 0.20\}$. For each sample, the parameters were estimated using three methods: L-moments, Maximum Likelihood Estimation (MLE), and HMC with a Beta prior for the shape parameter. The two frequentist estimators (L-moments and MLE) were obtained using the \texttt{extRemes} package \citep{Gilleland2016extRemes} in \texttt{R}. The Bayesian estimates were produced by using our custom library \texttt{extremMHMC}.

For each combination of $n$ and $\xi$, Bayesian estimation is performed using four independent Markov chains, each run for 4000 iterations with a burn-in period of 1000. For all GEV parameters, convergence is assessed using the rank-normalized split $\hat{R}$, with the threshold $\hat{R} \le 1.01$. In addition, we calculate both bulk-ESS and tail-ESS with a threshold $\ge 400$ to ensure reliable posterior estimation. Figure~\ref{fig:sim_results} shows that the simulation varies across different sample sizes and shape parameters.

In Figure~\ref{fig:sim_results}, we observe that the Beta prior induces a slight shrinkage of the estimated shape parameter towards the prior mean, particularly for the smallest sample size ($n = 15$) and for shape parameters $\xi \in \{-0.05, 0.20\}$. This results in a some bias. In the context of engineering this design, this bias is more of a feature than a bug of the algorithm. Engineering reliability analysis, relies on severe extrapolations of from the tail of the ground snow load distribution (see \cite{Bean2021SnowLoad} and \cite{liel2017} for examples). The bias helps to drastically reduce the variance in the estimates of $\xi$, which regularizes unrealistic tail behavior and produces more robust RTSL estimates. This reduction is variance is observed in the reduced values of RMSE, relative to frequentest approaches as shown in Table~\ref{tab:RMSE_simstudy}. For parameter $\theta$ and $l$ replications RMSE is defined as follows:

\begin{equation}\label{eq:rmse}
\text{RMSE}_\theta = \sqrt{ \frac{1}{l} \sum_{i=1}^l (\hat{\theta}_i - \theta)^2}
\end{equation}

Table~\ref{tab:RMSE_simstudy} illustrates that for all combinations of the shape parameter ($\xi$) and sample size ($n$), the HMC algorithm with a Beta prior yields consistently smaller RMSE values than both MLE and L-moments. The relative RMSE reduction achieved by HMC is always greater than 40\% compared to MLE and greater than 30\% compared to L-moments. Also, for small sample sizes ($n = 15$ and $n = 25$) and for positive parameters, RMSE reductions always exceed 50\% relative to both frequentist estimators. Although the relative percentage gap narrows as the sample size increases, HMC still exhibits a substantial reduction in RMSE compared to both frequentist approaches. These results highlight the practical advantage of the proposed Bayesian framework for estimating the GEV shape parameter compared to the frequentist methods.

\section{Application}\label{section:application}

\subsection{Data}

We use the preprocessed dataset from "The 2020 National Snow Load Study" by \cite{Bean2021SnowLoad} for this study which contains annual maximum snow load data for 9,715 stations, referred to hereafter as NSLS2020. This was developed using data from the National Oceanic and Atmospheric Administration's (NOAA) GHCN-Daily records of water-equivalent snow on the ground (WESD) and snow depth (SNWD) \citep{Menne2012anoverview}. The raw download included over 237 million observations from over 65,000 stations in the United States and Canada. Before refitting, NSLS2020 utilized both GHCN quality flags (QFLAG) and additional automated and targeted manual outlier screening, based on \cite{Durre2010comprehensive} (e.g., implausible unit jumps, WESD or SNWD inconsistencies, and anomalously heavy tails). It also defined snow seasons as October to June and created annual (seasonal) maxima. NSLS2020 outlines a clustering approach to reduce spatial redundancy in station records. Here, we retained the full station-level dataset and conducted distribution fitting prior to applying any clustering.

\subsection{Model Fitting and Results}

\begin{figure}[htbp]
    \centering
    \includegraphics[width=25pc, height = 25pc]{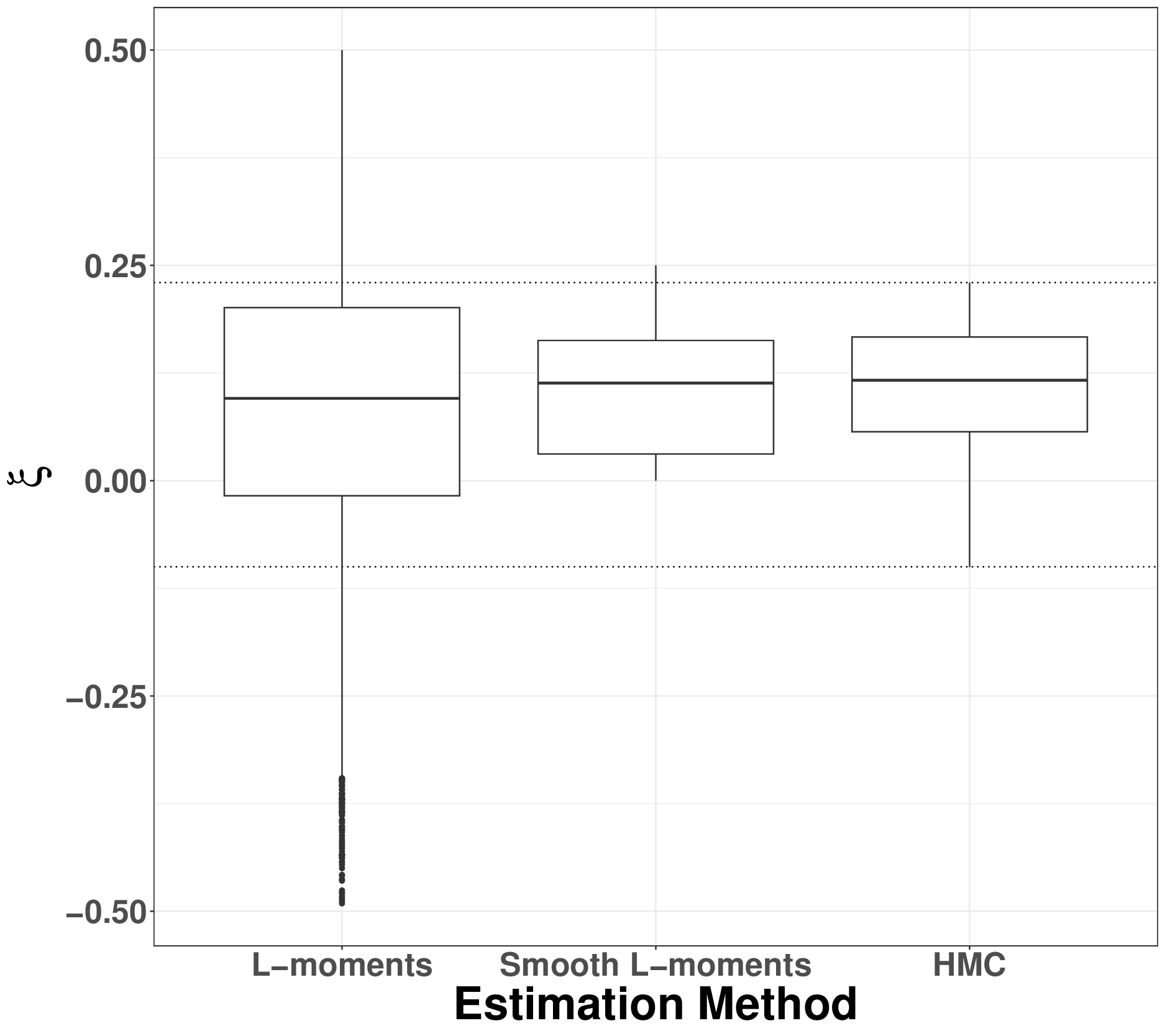}
    \caption{Boxplots summarizing the estimated GEV shape parameter $\xi$ obtained using L-moments, Smooth L-moments, and HMC with informative priors. The dotted horizontal lines denote the defined range $\xi \in [-0.1,\,0.23]$.}
    \label{fig:comp_methods}
\end{figure}

First, we fit the HMC with the priors defined in Equation~\ref{eq:priors} and identified 480 ($\approx 5\%$ out of 9715 stations) stations that fall outside our defined interval $[-0.1, 0.23]$. Then, as the second step, we refit the HMC using the priors defined in Equation~\ref{eq:priorsnew} for those 480 stations to ensure the parameter remains within the defined range. In both steps, we used four independent chains with 3000 iterations and a 1000 burn-in period per chain. Also, to assess convergence and sampling efficiency, used the rank-normalized split $\hat{R}$, as well as bulk-ESS and tail-ESS. For all 9715 measurement locations, $\hat{R}$ remained less than 1.01, and both bulk-ESS and tail-ESS were greater than 400 for all GEV parameters independently. Comparison for shape parameter estimated using L-moments, Smooth L-moments \citep{Bean2021SnowLoad} and HMC methods shown in Figure \ref{fig:comp_methods} and we can see that the standard L-moments distributed whole domain of the shape parameter $[-0.5,0.5]$. The Standard L-moments shape parameter was spatially smoothed using a Generalized Additive Model (GAM) as a function of the median of annual maxima to obtain the smooth L-moments for the shape parameter \citep{Bean2021SnowLoad}. The resulting smoothed values were truncated to lie within the interval  $[0,\, 0.25]$, meaning values below $0$ were set to $0$, and values above $0.25$ were capped at $0.25$. Here, HMC estimates have a slightly higher median and lower variability than Smooth L-moments estimates for the across all 9715 station shape parameters.

Several key properties motivate the use of the Bayesian HMC approach. One main advantage of this method is that it naturally propagates uncertainty without letting that uncertainty overwhelm the RTSL estimation process. Also, instead of regularizing the point estimate of the $\xi$ parameter (tail behavior of the distribution) spatially, as in the smooth L-moments approach, HMC regularizes it in a probabilistically coherent way without post hoc constraints. Since the GEV $\xi$ is well known to be unstable under stations with short record lengths and in the presence of extreme winters, incorporating geophysically motivated prior in the HMC framework provides another advantage to directly stabilize inference on $\xi$. This avoids estimating impractical $\xi$ parameters that can inflate RTSL and limit the influence of high-leverage years. Additionally, spatial smoothing (as in the smooth L-moments) can bias the shape parameter estimates toward those at neighboring locations, potentially inconsistent with local climatology or topography. In contrast, the HMC method is primarily driven by individual-station data and therefore avoids any unintended consequences of spatial smoothing.

To compare RTSL estimates for Risk category II (reliability index $\beta$ =  3.0) across three methods, we return to the four snow measurements locations referred to in the simulation study. These cities chosen because they reflect the different types of extreme snow load distributions represented in the larger dataset. 

\begin{itemize}

     \item \textbf{Denver, CO:} Denver snow patterns follow a heavy-tailed distribution due to infrequent but exceptionally intense snowstorms from orographic lifting by the Rocky Mountains.

     \item \textbf{Washington, DC:} Snow load patterns in Washington, DC, generally exhibit a heavy-tailed distribution, as occasional high-impact snowstorms can produce unusually large annual maxima.
     
     \item \textbf{Duluth, MN:} Duluth snow patterns follow a light-tailed distribution due to frequent moderate snowstorms rather than rare, extreme events, resulting in less variability in annual maximum snow loads and fewer extreme outliers.

     \item \textbf{Sierra City, CA:} Sierra City is located in the heavily snow-affected Sierra Nevada region, which experiences frequent snowfall. Sometimes, this region receives several feet of snow in a single storm. This suggests that Sierra City exhibits a light-tailed snow load distribution.

\end{itemize}

\begin{table}[htbp]
\caption{GEV parameter estimates and RTSL values for all selected cities (Denver, CO; Washington, DC; Duluth, MN; Sierra City, CA) based on three estimation methods. LM, SM-LM, and HMC represent estimates obtained from L-moments, Spatially Smoothed L-moments, and Hamiltonian Monte Carlo with a Beta prior distribution, respectively.}
\label{tab:RTL_comparison}
\centering

\begin{tabular*}{\textwidth}{@{\extracolsep{\fill}}c c c c c c c c@{}}
\hline
Station &
\begin{tabular}{@{}c@{}}Year \\ Range\end{tabular} &
\begin{tabular}{@{}c@{}}Sample \\ Size\end{tabular} &
Method &
$\mu$ & $\sigma$ & $\xi$ &
\begin{tabular}{@{}c@{}}RTSL \\ (kPa)\end{tabular} \\
\hline

\multirow{3}{*}{Denver} &
\multirow{3}{*}{1948-2020} &
\multirow{3}{*}{73} &
LM  & 0.199 & 0.094 & 0.256 & 2.71 \\
& & & SM-LM  & 0.200 & 0.094 & 0.246 & 2.55 \\
& & & HMC & 0.202 & 0.097 & 0.212 & 2.18 \\
\hline

\multirow{3}{*}{Washington} &
\multirow{3}{*}{1946-2020} &
\multirow{3}{*}{63} &
LM  & 0.199 & 0.136 & 0.232 & 3.27 \\
& & & SM-LM  & 0.198 & 0.129 & 0.250 & 3.46 \\
& & & HMC & 0.202 & 0.136 & 0.211 & 2.87 \\
\hline

\multirow{3}{*}{Duluth} &
\multirow{3}{*}{1948-2020} &
\multirow{3}{*}{73} &
LM & 1.006 & 0.456 & -0.163 & 3.27 \\
& & & SM-LM  & 0.970 & 0.427 & 0.000 & 4.36 \\
& & & HMC  & 0.980 & 0.442 & -0.051 & 3.94 \\
\hline

\multirow{3}{*}{Sierra City} &
\multirow{3}{*}{1915-1994} &
\multirow{3}{*}{39} &
LM  & 2.258 & 1.576 & -0.090 & 11.81 \\
& & & SM-LM & 2.175 & 1.415 & 0.044 & 15.56 \\
& & & HMC & 2.194 & 1.494 & 0.012 & 14.77 \\
\hline

\end{tabular*}
\end{table}

\begin{figure}[htbp]
    \centering
    \includegraphics[width=30pc, height=30pc]{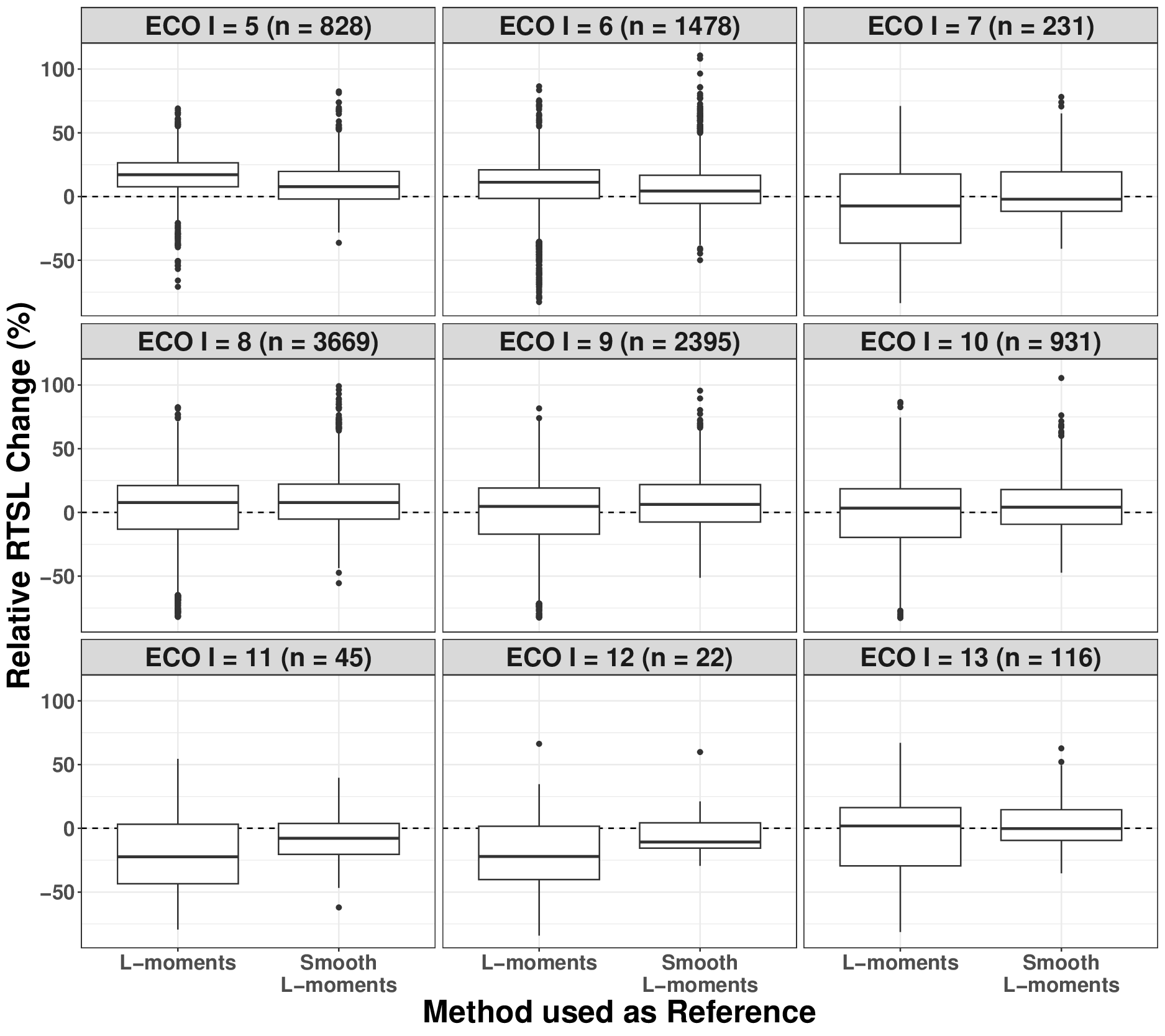}
    \caption{Boxplots of relative RTSL changes (\%) by Level 1 ecological region, computed using Equation~\ref{eq:rel_change}. The value $n$ denotes the number of stations within each region.}
    \label{fig:rel_change_box}
\end{figure}

As shown in Table~\ref{tab:RTL_comparison}, the standard L-moments estimated $\xi$ parameter for the Denver location shows an impractical $\xi$ value where $\xi > 0.25$. This implies that the estimated RTSL values are generally unacceptable based on the L-moments approach. On the other hand, the HMC-based RTSL value is consistently lower than the smooth L-moments-based RTSL value. This relative reduction in the HMC-based RTSL value compared to the smooth L-moments-based value is around 14.5\%. In Washington, DC, the standard L-moments estimated the $\xi$ parameter to be less than 0.25, and the smooth L-moments estimated the $\xi$ parameter equal to 0.25. This occurs because the spatial smoothing approach of \cite{Bean2021SnowLoad} caps the maximum value of the shape parameter at 0.25, thereby allowing smooth L-moments estimates to fall between 0.23 and 0.25. The relative reductions in HMC-based RTSL values are approximately 12\% and 17\% compared to the standard L-moment- and smooth L-moment-based RTSL values, respectively.

In Duluth, the HMC method-based RTSL value is generally higher than those derived from standard L-moments. This reflects the sensitivity of design loads to small changes in the estimate of the $\xi$ parameter in light-tail regions. For the standard L-moments estimates in Duluth, the $\xi$ value lies outside our geophysically motivated shape parameter range $[-0.1, 0.23]$. The smooth L-moments shape parameter equals 0.0, which is due to the lower bound of $\xi = 0$ imposed in the spatial smoothing framework of \cite{Bean2021SnowLoad}. The HMC-based RTSL value decreased by approximately 9\% compared to the smooth L-moments-based estimate. For Sierra City, only the standard L-moments estimated $\xi$ parameter value is negative, but it remains within the range $[-0.1, 0.23]$. Because of this, the standard L-moments-based RTSL value is lower than both the smooth L-moments-based RTSL value and the HMC-based RTSL value. Additionally, the HMC-based RTSL value is lower than the smooth L-moments-based RTSL value by around 5\%.

\begin{figure}[htbp]
    \centering
    \includegraphics[width=40pc, height=40pc]{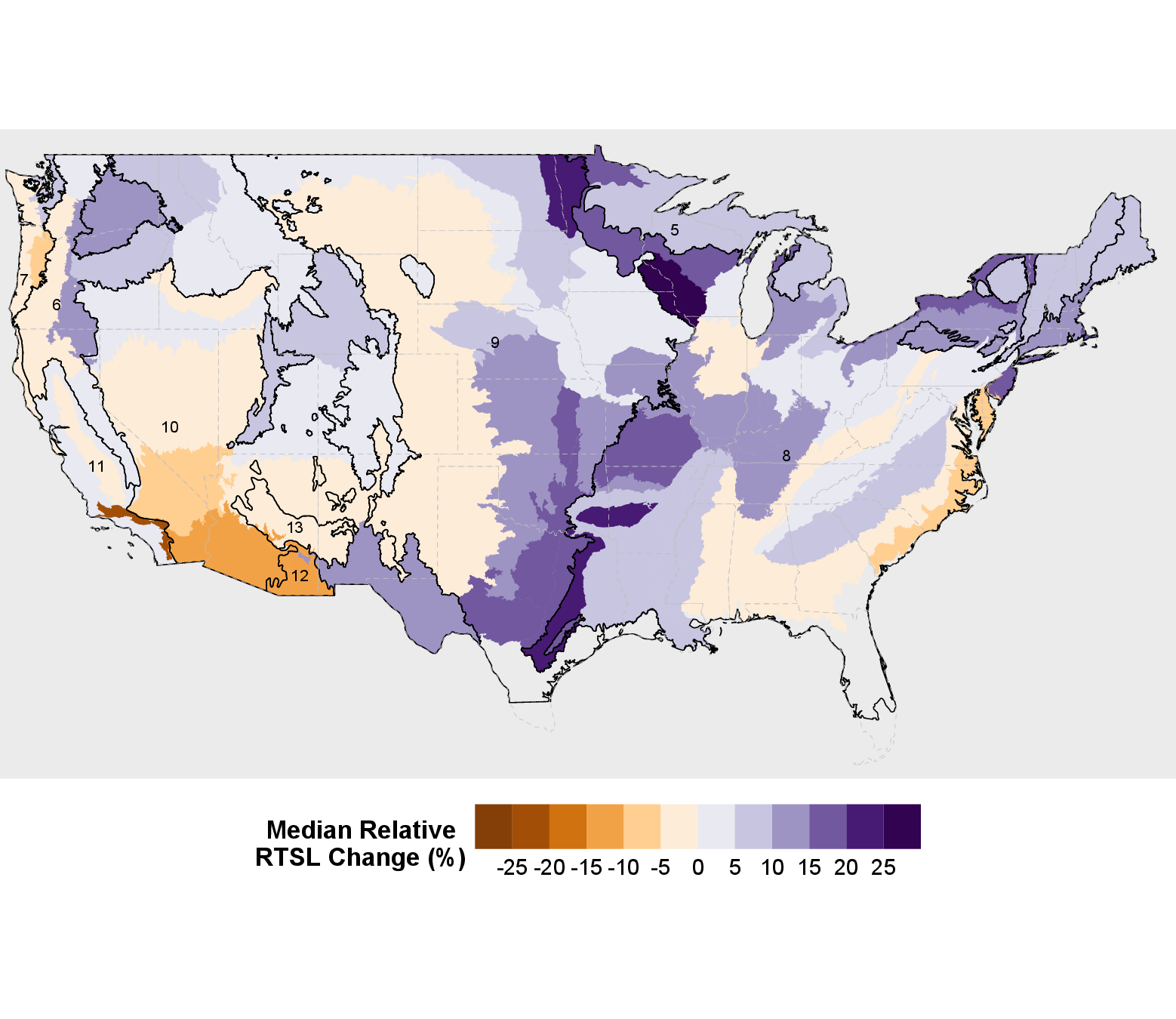}
    \caption{The median relative percentage changes in Reliability-Targeted Snow Loads (RTSL) relative to the smooth L-moment-based estimates across the Conterminous United States (CONUS). Positive percentages indicate higher HMC-based RTSL values compared to the smooth L-moments based RTSL values, while negative values indicate lower values.}
    \label{fig:rel_change_conus}
\end{figure}

To provide a national-level summary of these relative RTSL value changes compared to standard L-moments based values and smooth L-moments based values, we construct level 1 Ecological region-wise boxplots across all stations in the NSLS2020 dataset. The ecological regions considered in this study correspond to Level 1 regions 5-13 as defined by \citep{CEC1997}, namely Northern Forests (5), Northwestern Forested Mountains (6), Marine West Coast Forests (7), Eastern Temperate Forests (8), Great Plains (9), North American Deserts (10), Mediterranean California (11), Southern Semi-Arid Highlands (12), and Temperate Sierras (13). To calculate relative changes as percentages, we used equation~\ref{eq:rel_change}, and the boxplots are shown in Figure~\ref{fig:rel_change_box}. 

\begin{equation}\label{eq:rel_change}
\text{Relative RTSL Change} (\%) = \frac{\text{RTSL}_{HMC} - \text{RTSL}_{Ref}}{\text{RTSL}_{Ref}} \times 100
\end{equation}
Here, $\text{RTSL}_{Ref}$ represents the RTSL values from either the standard L-moment based method or the smooth L-moments based method.

Relative changes in HMC-based RTSL values compared to standard L-moment-based values for ECO regions 5, 6, 8, 9, and 10 are generally modest, with the central bulk of relative percentage changes lying within about $\pm 25\%$. The median relative percentages for these regions are approximately 17\%, 11\%, 7.5\%, 4.5\%, 3.5\%, respectively. This suggests that HMC-based values tend to yield slightly higher RTSLs in these regions compared to standard L-moment-based values. In contrast, ECO regions 7, 11, 12, and 13 show substantially greater variability in relative changes (predominantly reductions), with median relative percentage changes of approximately -7\%, -22\%, -22\%, and 2\%, respectively. One plausible reason for the increased variability is the smaller number of stations in these regions (especially ECO 11 and 12) compared to the larger-sample regions.

Compared to smooth L-moment-based RTSL values, relative changes in HMC-based values follow the same pattern as with the standard L-moment-based values, with a similar or slightly lower variability. In all regions central bulk of relative percentage changes lies within about $\pm 25\%$. For ECO regions 5, 6, 8, 9, and 10, the median relative changes are approximately 7.5\%, 4\%, 7.5\%, 6\% and 4\%, respectively. Conversely,  ECO regions 7, 11, 12, and 13 exhibit negative median relative changes of approximately -2\%, -7.5\%, -11\% and -0.2\%, respectively. The slightly reduced variability relative to the standard L-moment comparison is consistent with smoothing and post-hoc constraints, which moderates station-to-station fluctuations in GEV shape estimates and thereby reduces the magnitude of method-to-method differences. Although most stations fall within $\pm 25\%$, several regions contain notable outliers (around 1.5\% stations), indicating that a small subset of stations experiences much larger departures between methods. Figure~\ref{fig:rel_change_conus} presents the median relative percentage changes relative to the smooth L-moments for the Conterminous United States (CONUS), including the within-region variability.  The median relative percentage changes exhibit substantial regional variability, with predominantly positive changes in parts of the Upper Midwest, Central, and Southern regions, and negative changes concentrated in portions of the Southwest, West Coast, and East Coast.

\section{Conclusion}\label{section:conclusion}

In this article, we developed a Bayesian framework to estimate the parameters of the Generalized Extreme Value distribution for annual snow load extremes using the Hamiltonian Monte Carlo (HMC) with the Dual Averaging (DA) algorithm. Our primary goal was to obtain a stable estimate of the shape parameter ($\xi$) that yields realistic reliability-targeted snow load estimates (RTSLs). Estimating $\xi$ was critical because it largely governs the estimation of super-extreme snow load quantiles as part of an RTSL analysis. To identify a plausible interval for the shape parameter, we conducted a literature review focused on extreme precipitation and rainfall events, given the limited number of studies on extreme snow load events, and as snow is a fundamentally precipitation-driven process. During the literature review, we identified a reasonable range for $\xi$, which is $[-0.1,0.23]$. The next step was to identify a reasonable prior distribution for the $\xi$. We adopted a Beta distribution with support $[-0.5, 0.5]$ as the prior distribution, which was motivated by prior work on extreme hydrological events. Since the $\xi$ parameter range is primarily positive but could also be negative (mostly at high elevations), we needed prior distribution parameters to account for both cases. Our approach was intended to be a global approach, as the prior parameters were defined using the hyperbolic tangent function and the natural logarithm of annual maxima. This helped to smooth the transition between the positive and negative regions of the interval, depending on the data. 

With this prior distribution, we conducted a simulation study across different combinations of shape parameters and sample sizes, with 1000 replications per combination, and compared the resulting RMSE values from two frequentist approaches (L-moments and MLE) with those from our Bayesian framework (HMC). The RMSE values showed a significant reduction for the HMC method compared to the two frequentist methods. After the simulation study, we estimated the GEV parameters and RTSLs for the real-world NSLS2020 dataset, which contained annual maximum snow load records for 9715 weather stations. For the stations where the parameter lay outside the desired interval $[-0.1, 0.23]$, we refit the model using Normal priors with a small variance to ensure the estimates remained within physically plausible bounds. Additionally, we compared all parameter estimates and RTSLs with those obtained using L-moments and spatially smoothed L-moments.

We then compared the relative changes in RTSL values derived from HMC-based parameter estimates with those obtained from both the standard L-moment and smoothed L-moment methods across Level 1 ecological regions in the United States. The relative changes for HMC versus standard L-moment RTSL values exhibited greater variability in some regions compared to the HMC versus smoothed L-moment comparison. However, across all ecological regions, the central bulk of relative percentage changes lay within approximately $\pm 25\%$ for the HMC versus smoothed L-moments RTSL comparison.

Although the Bayesian HMC approach is robust, it has several limitations. One key limitation is that the estimated parameters may exhibit slight bias towards the prior mean. This may be a problem is the mean of the prior distribution is severely misspecified, which we have tried to avoid in this study by allowing the prior distribution center to shift based on local snow accumulation dynamics. Another limitation is that, unlike frequentist methods, the Bayesian approach is computationally expensive. Additionally, the estimated parameters rely entirely on the assumption that the original data are correct meaning that annual maximum series are stationary and free from systematic measurement errors. Future work could address these limitations by developing a hierarchical Bayesian framework that directly accounts for spatial constraints within the model. This can be done by allowing the Beta prior parameters to vary spatially using a Gaussian process within a hierarchical framework. This would enable nearby stations to share information while maintaining physically plausible bounds on the shape parameter. Additional extensions may include the incorporation of nonstationary GEV models with other climate variables such as temperature and computational optimizations for large-scale implementation.

However, our Bayesian approach provides a statistically coherent and physically grounded framework for shape parameter estimation. This method stabilizes tail inference without using additional spatial smoothing and post-hoc constraints. In addition, this framework naturally propagates uncertainty when estimating the shape parameter and preserves local station behavior. Because of these characteristics, this Bayesian method provides a significant advantage over the currently existing smooth L-moments method, especially in situations with short records and highly fluctuating extremes.

\section*{Author contributions}

\textbf{Shaveen Britto:} Conceptualization; methodology; investigation; formal analysis; software; visualization; writing - original draft; writing - review and editing. \textbf{Brennan Bean:} Conceptualization; methodology; data curation; supervision; writing - review and editing.

\section*{Acknowledgments}
The authors wish to thank Nicholas Brimhall, an undergraduate student at Utah State University, for guidance in using the international database of in-situ snow load measurements described in this paper. 

All analysis for this paper was performed in R 4.5.0 \citep{Rlang} with the following packages:

\begin{itemize}
    \item rstan~\citep{rstan}
    \item extRemes~\citep{Gilleland2016extRemes}
    \item snowload2~\citep{Bean2021SnowLoad}
    \item Rcpp~\citep{Eddelbuettel2011rcpp}
    \item parallel~\citep{Rlang}
    \item tidyverse~\citep{Wickham2019tidyverse}
    \item sf~\citep{Pebesma2018sf}
    \item rnaturalearth~\citep{Massicotte2026rnaturalearth}
    \item RColorBrewer~\citep{Neuwirth2022RColorBrewer}
\end{itemize}

\section*{Financial disclosure}

This study was supported by the National Oceanic and Atmospheric Administration (NOAA) Modeling, Analysis, Predictions, and Projections (MAPP) Program under Grant No. NA23OAR310609.

\section*{Conflict of interest}

The authors declare no potential conflict of interests.

\bibliographystyle{plainnat}
\bibliography{references}

\appendix

\section{Supplementary Details}
\vspace*{12pt}

For Appendix~\ref{appendix:A1} and Appendix~\ref{appendix:A2}, the details and equations are drawn from the studies of \cite{Cannas2003introduction}, \cite{Neal2011mcmc}, \cite{Thomas2020learning}, \cite{Betancourt2017aconceptual}, \cite{Beskos2013optimal}, \cite{Hoffman2014thenuts}, \cite{Nesterov2009primal}, \cite{Geyer1992practical}, and \cite{Vehtari2021rank}. For Appendix~\ref{appendix:A3} and Appendix~\ref{appendix:A4}, we derived the analytical gradients for the HMC algorithm and developed the \texttt{extremMHMC} package in \texttt{R}. Equations from these references are repeated in this appendix for convenience in review.

\subsection{Review of HMC algorithm\label{appendix:A1}}
\vspace*{12pt}

In physics, the Hamiltonian function ($H(\boldsymbol{q},\boldsymbol{p})$) describes the total energy of a system using the current position or generalized coordinate of the particle or object ($\boldsymbol{q}$) and the generalized momentum of that object ($\boldsymbol{p}$). The Hamiltonian function has two key terms. One is "Potential energy" ($U(\boldsymbol{q})$) and the other one is "Kinetic energy" ($K(\boldsymbol{p}$). In physics, these two energy terms together describe the total energy of the system, and total energy is always constant for a frictionless surface. For dimension $d$, we can write the Hamiltonian equation as,
\begin{equation}\label{eq:standardhmfunction}
H(\boldsymbol{q},\boldsymbol{p}) = U(\boldsymbol{q}) + K(\boldsymbol{p}) \quad \text{where} \quad \boldsymbol{q},\boldsymbol{p} \in \mathbb{R}^d
\end{equation}
Here, Potential energy depends only on the current position of the object ($\boldsymbol{q}$) and Kinetic energy depends only on the momentum of the object ($\boldsymbol{p}$). Further technical and mathematical details see \cite{Cannas2003introduction}.

\begin{figure}[htbp]
\centering
\includegraphics[width=\textwidth]{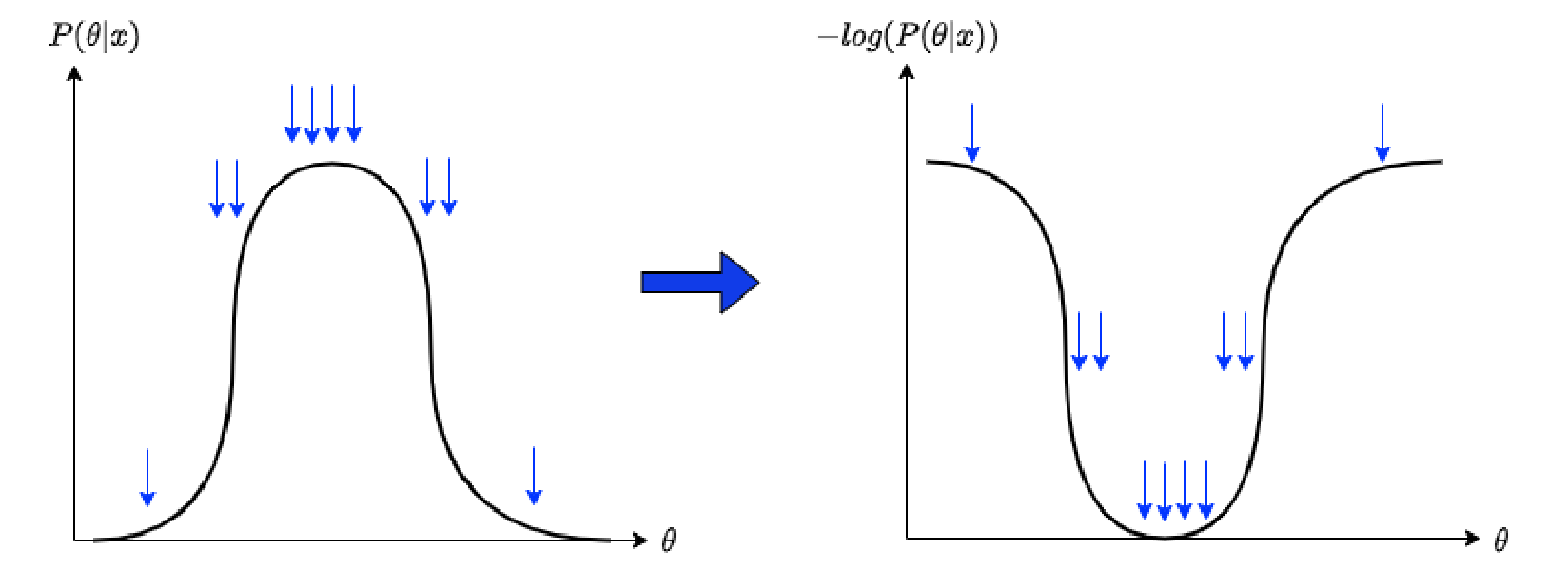} 
\caption{Approximate relationship between Posterior density functions and $\theta$.}
\label{fig:posteriorrel}
\end{figure}

In a probabilistic setting, consider a univariate probability distribution with parameter $\theta$ and for given data vector $\boldsymbol{x}$. Define the posterior density function as $P(\theta|\boldsymbol{x})$, which has a bell-shaped curve. Figure~\ref{fig:posteriorrel} shows the approximate relationship between $P(\theta|\boldsymbol{x})$ versus $\theta$ and $-\log(P(\theta|\boldsymbol{x}))$ versus $\theta$. According to the Figure~\ref{fig:posteriorrel}, generating samples from the bottom of the right-hand $-\log(P(\theta|\boldsymbol{x}))$ implies that we are generating samples from the higher density regions of the left-hand side $P(\theta|\boldsymbol{x})$.

Motivated by this relationship, in standard HMC, $\boldsymbol{\theta}$ sampling from the posterior density function ($\boldsymbol{\theta} \sim P(\boldsymbol{\theta}|\boldsymbol{x})$) and auxiliary momentum variables generated from the Multivariate Normal Distribution with mean vector with zeros and a user defined covariance matrix ($\boldsymbol{M}$) ($\boldsymbol{p} \sim \mathcal{N}_d(\mathbf{0}, \boldsymbol{M})$). Usually, this $\boldsymbol{M}$ is a scalar multiple Identity matrix \citep{Neal2011mcmc}. Then in HMC, $U(\boldsymbol{\theta})$ define as $-\log(P(\boldsymbol{\theta}|\boldsymbol{x}))$ and $K(\boldsymbol{p})$ as $\frac{1}{2}\boldsymbol{p}^T \mathbf{M}^{-1}\boldsymbol{p}$. Here, $\boldsymbol{\theta}$ and $\boldsymbol{p}$ has independent canonical distributions. Then we can write the Hamiltonian function as follows,

\begin{equation}\label{eq:hmcfunction}
H(\boldsymbol{\theta},\boldsymbol{p}) = -\log(P(\boldsymbol{\theta}|\boldsymbol{x})) + \frac{1}{2}\boldsymbol{p}^T \mathbf{M}^{-1}\boldsymbol{p} \quad \text{where} \quad \boldsymbol{\theta},\boldsymbol{p} \in \mathbb{R}^d
\end{equation}
Also, we can write Hamiltonian equations as follows,

\begin{equation}\label{eq:gradhmcfunction}
\begin{aligned}
\frac{d\boldsymbol{\theta}}{dt} &= \mathbf{M}^{-1}\boldsymbol{p} \\ 
\frac{d \boldsymbol{p}}{dt} &= \nabla_{\boldsymbol{\theta}} \log(P(\boldsymbol{\theta}|\boldsymbol{x}))    
\end{aligned}
\end{equation}

Where $\nabla_{\boldsymbol{\theta}} log(P(\boldsymbol{\theta}|\boldsymbol{x}))$ is the gradient of the log posterior density function. The idea is that by introducing auxiliary variables $\boldsymbol{p}$ and incorporating gradient information, the exploration of the parameter space becomes more efficient. It is often difficult to solve the Hamiltonian equations analytically, so numerical methods such as the \textit{Euler} method were initially used to solve these equations. However, after \cite{Neal2011mcmc} identified errors with these methods, the \textit{leapfrog} method (a modified Euler method) is now commonly used to solve Hamiltonian equations in practice. For user defined step-size ($\epsilon$), the \textit{leapfrog} algorithm proceeds as follows,

\begin{equation}\label{eq:leapfrog}
\begin{aligned}
\boldsymbol{p}^{t + \frac{\epsilon}{2}} &= \boldsymbol{p}^t + \frac{\epsilon}{2}\nabla_{\boldsymbol{\theta}} \log(P(\boldsymbol{\theta}|\boldsymbol{x}))  \\ 
\boldsymbol{\theta}^{t + \epsilon} &= \boldsymbol{\theta}^t + \epsilon\mathbf{M}^{-1}\boldsymbol{p} \\
\boldsymbol{p}^{t + \epsilon} &= \boldsymbol{p}^{t + \frac{\epsilon}{2}} + \frac{\epsilon}{2}\nabla_{\boldsymbol{\theta}} \log(P(\boldsymbol{\theta}|\boldsymbol{x}))
\end{aligned}    
\end{equation}

After performing multiple \textit{leapfrog} steps ($L$), we obtain the proposed parameters ($\boldsymbol{\theta}^*$) for the posterior density function \citep{Thomas2020learning}. An important study by \cite{Betancourt2017aconceptual} shows that, due to its Symplectic property (i.e, it preserves the area or volume in phase space during the integration), the \textit{leapfrog} algorithm preserves the fundamental geometric structure (such as energy and momentum) of Hamiltonian systems, resulting in superior long-term accuracy and stability in simulations. This study also shows that, while the HMC algorithm predominantly samples from regions of higher posterior density, it still effectively samples the tail regions of the distribution as well. Sampling these tail regions is crucial in our study because the accurate estimation of the 'shape' parameter in the GEV distribution depends on effectively capturing the tail behavior. Failure to adequately sample the tails can lead to biased or imprecise parameter estimates, which in turn affect the reliability of extreme value predictions. Even though HMC gives a high acceptance probability because of the Symplectic property, several studies mentioned that tuning $L$ and $\epsilon$ to maintain the acceptance rate between 65\% and 80\% helps to balance the computational cost and acceptance to maximize efficiency \citep{Beskos2013optimal, Hoffman2014thenuts}. For more theoretical aspects of the HMC algorithm, see \cite{Neal2011mcmc} and \cite{Betancourt2017aconceptual}. 

Even though HMC handles the complex geometry of a distribution properly, it has a significant disadvantage: it is considerably slower than the M-H algorithm because both $L$ and $\epsilon$ must be manually tuned to achieve good convergence of the posterior parameters. To overcome this problem  \cite{Hoffman2014thenuts} introduced a new approach and a more developed version for HMC called "Dual Averaging (DA)"  introduced by \cite{Nesterov2009primal} to tune the $\epsilon$ for the best value using the few initial iterations (called warm-up phase or adaptation phase) of the HMC algorithm. This DA algorithm is designed to find the optimal solution to non-smooth, stochastic convex optimization problems using subgradients. This DA algorithm provides automatic, stable, and robust adaptation of the $\epsilon$. Additionally, dual averaging includes an implicit form of regularization that decays over iterations, improving convergence stability and preventing overly aggressive step-size parameter changes early in the adaptation phase. The developed version of DA used in HMC is described in Algorithm 5 of \cite{Hoffman2014thenuts}. Once we have the $\epsilon$, we can define trajectory length ($\lambda$) where $\lambda = \epsilon L$ to find $L$.

\subsection{Calculating Potential Scale Reduction Factor and Effective Sample Size\label{appendix:A2}}
\vspace*{12pt}
For single chain ESS estimate as follows:

\begin{equation}\label{eq:Ness_inf}
\begin{aligned}
N_\text{ESS} = \frac{N}{1 + 2\sum_{t = 1}^\infty \rho_t}
\end{aligned}    
\end{equation}

Where $N$ is the number of posterior samples after burn-in phase and $\rho_t$ is autocorrelation at lag $t \geq 0$. Since always we have finite length chains, \cite{Geyer1992practical} truncated the infinite sum to some finite lag $T$.

\begin{equation}\label{eq:Ness_T}
\begin{aligned}
N_\text{ESS} = \frac{N}{1 + 2\sum_{t = 1}^T \hat{\rho}_t}
\end{aligned}    
\end{equation}

Selecting $T$ is crucial and finding $T$ as follows:
\begin{enumerate}[1.]
    \item Compute $\hat{\rho}_t$ for $t = 1,2,3, \hdots$.
    \item Compute paired sums $\hat{P}_{t'} = \hat{\rho}_{2t'} + \hat{\rho}_{2t' + 1}$ for $t' = 0, 1,2,3, \hdots$.
    \item Select the largest $k$ such that all $\hat{P}_{t'} > 0$ for $t' = 0, 1,2,3, \hdots, k$ (i.e., continue computing $\hat{P}_{t'}$ until the first non–positive value)
    \item Set $T = 2k + 1$
\end{enumerate}
\cite{Geyer1992practical} use this truncation because at large lags, sample $\hat{\rho}_t$ are mostly noise and when the paired sum ($\hat{P}_{t'}$) first goes non-positive, indicates the beginning of this noise region and provides an appropriate point at which to truncate the autocorrelation series. \cite{Vehtari2021rank} extended the ESS calculation approach from single chain to multi chain approach to get a more robust estimation for ESS. To calculate $\hat{\rho}_t$ for $K$ number of multiple independent chains as follows:

\begin{equation}\label{eq:ess}
\begin{aligned}
\hat{\rho}_t &= 1 - \frac{W - \frac{1}{K} \sum_{k = 1}^K W_k \hat{\rho}_t^{(k)}}{\hat{V} }
\end{aligned}
\end{equation}

Where $W_k$ is the Within-chain variance of chain $K$ and $W = \frac{\sum^K_{i = 1} W_i}{K}$.

For example if we have $K$ chains for univariate parameter $\theta$ with $L$ number of posterior samples per chain, first combined those samples and assign ranks $r^{(kl)}$ for each sample. Then calculate the rank to normal scores as follows:

\begin{equation}\label{eq:ranknormalize}
\Tilde{\theta}^{(kl)} = \Phi^{-1} \left( \frac{r^{(kl)} - \frac{3}{8}}{KL - \frac{1}{4}} \right)
\end{equation}

Similarly, we obtain $\hat{R}_2$ by first folding the rank-normalized draws around the median, 
$\tilde{\theta}^{(kl)}_{\text{fold}} = \bigl|\tilde{\theta}^{(kl)} - \operatorname{median}(\tilde{\theta})\bigr|$, 
and then recomputing the split-\mbox{$\hat{R}$} on these folded values, which makes $\hat{R}_2$ more sensitive to lack of convergence in the tails.

where $\Phi^{-1}(.)$ is the inverse CDF of the standard normal distribution. Then using $\Tilde{\theta}^{(kl)}$ calculate the $\hat{R}_1$. Similarly, they obtain $\hat{R}_2$ by first folding the rank-normalized draws around the median. 

\begin{equation}\label{eq:folding}
\tilde{\theta}^{(kl)}_{\text{fold}} = |\tilde{\theta}^{(kl)} - \operatorname{median}(\tilde{\theta})|
\end{equation}

and calculate $\hat{R}$ as,

\begin{equation}\label{eq:maxrhat}
\hat{R} = \max \{\hat{R}_1, \hat{R}_2 \}
\end{equation}

$\hat{R}_1$ is more sensitive to lack of convergence in the bulk part of the posterior distribution (central region), and $\hat{R}_2$ is more sensitive to discrepancies in the tail behavior of the posterior distribution. The final $\hat{R}$ determines that lack of convergence in either the bulk or the tails will be reflected in the posterior chain. Note that, before calculating $\Tilde{\theta}^{(kl)}$, divide each chain into two halves to increase the number of chains. Dividing a chain into two parts increases the sensitivity of the diagnostics to non-stationarity and helps to detect whether both halves converge to the same invariant parameters. 

Also, using $\Tilde{\theta}^{(kl)}$, divide the ESS calculation into two parts, namely Bulk-ESS, which calculates the ESS for the central region of the posterior distribution, and Tail-ESS, which calculates the ESS specifically for extreme quantiles (e.g., $5^\text{th}$ and $95^\text{th}$ percentiles) in the tails of the posterior. This central part of the posterior distribution is used to find the posterior estimate for the parameter (e.g., mean or median), and the tail part is used to find the quintiles (e.g., 95\% credible intervals) for the posterior estimates. So, this Bulk-ESS and Tail-ESS assess the accuracy of estimates separately.

\subsection{Analytical gradients for GEV log likelihood function\label{appendix:A3}}
\vspace*{12pt}
Let $z_i = \frac{x_i - \mu}{e^{\delta}}$ and $y_i = 1 + \xi z_i$

\begin{equation}\label{eq:gradmu}
\begin{aligned}
\frac{\partial \ell(\mu,\delta,\xi;\boldsymbol{x})}{\partial \mu} = \begin{cases}
\frac{1}{e^{\delta}}\left[ (1+\xi) \sum_{i=1}^n \frac{1}{y_i} - \sum_{i=1}^n y_i^{- \frac{1}{\xi} - 1} \right]  &, \xi \neq 0 \\
\frac{1}{e^{\delta}}\left[ n - \sum_{i=1}^n \exp \left(- z_i \right) \right]  &, \xi = 0
\end{cases}
\end{aligned}    
\end{equation}

\begin{equation}\label{eq:graddelta}
\begin{aligned}
\frac{\partial\ell(\mu,\delta,\xi;\boldsymbol{x})}{\partial \delta} = \begin{cases}
(1 + \xi) \sum_{i=1}^n \frac{z_i}{y_i} - n - \sum_{i=1}^n y_i^{-\frac{1}{\xi} - 1} z_i  &, \xi \neq 0 \\
\sum_{i=1}^n z_i - n -  \sum_{i=1}^n \exp\left(- z_i\right) z_i  &, \xi = 0
\end{cases}
\end{aligned}    
\end{equation}

\begin{equation}\label{eq:gradxi}
\begin{aligned}
\frac{\partial \ell(\mu,\delta,\xi;\boldsymbol{x})}{\partial \xi} =  \frac{1}{\xi^2} \sum_{i=1}^n \log(y_i) -\left(1+\frac{1}{\xi}\right) \sum_{i=1}^n \frac{z_i}{y_i} - \frac{1}{\xi^2} \sum_{i=1}^n y_i^{-\frac{1}{\xi}} \log(y_i)  + \frac{1}{\xi} \sum_{i=1}^n y_i^{-\frac{1}{\xi}-1} z_i, \quad \xi \neq 0
\end{aligned}    
\end{equation}

\subsection{Analytical gradients for log Priors\label{appendix:A4}}
\vspace*{12pt}
\begin{itemize}
    \item If $\theta \sim \mathcal{N}(\gamma_{\theta}, \tau_{\theta}^2) \ ; \quad -\infty < \gamma_{\theta} < \infty, \quad \tau_{\theta}^2 > 0$
    \begin{equation}\label{eq:gradnormal}
    \begin{aligned}
    \frac{\partial \log(f_{\Theta}(\theta))}{\partial \theta} = -\frac{\theta - \gamma_{\theta}}{\tau_{\theta}^2}
    \end{aligned}    
    \end{equation}

    \item If $\theta \sim \textit{Beta}(\eta_{\theta}, \kappa_{\theta}) \ ; \quad -0.5< \theta <  0.5, \quad  \eta_{\theta} > 0, \quad \kappa_{\theta} > 0$
    \begin{equation}\label{eq:gradbeta}
    \begin{aligned}
    \frac{\partial \log(f_{\Theta}(\theta))}{\partial \theta} = \begin{cases} 
    \frac{(\eta_{\theta} - 1)}{\theta + 0.5} - \frac{(\kappa_{\theta} - 1)}{0.5 - \theta} &, -0.5 < \theta < 0.5 \\
    0 &, o/w
    \end{cases}
    \end{aligned}    
    \end{equation}
\end{itemize}

\subsection{Frequency Distribution of Shape Parameters Outside [-0.1,0.23]\label{appendix:A5}}
\vspace*{12pt}

\begin{figure}[htbp]
    \centering
    \begin{subfigure}[b]{0.475\textwidth}
        \centering
            \includegraphics[width=\linewidth]{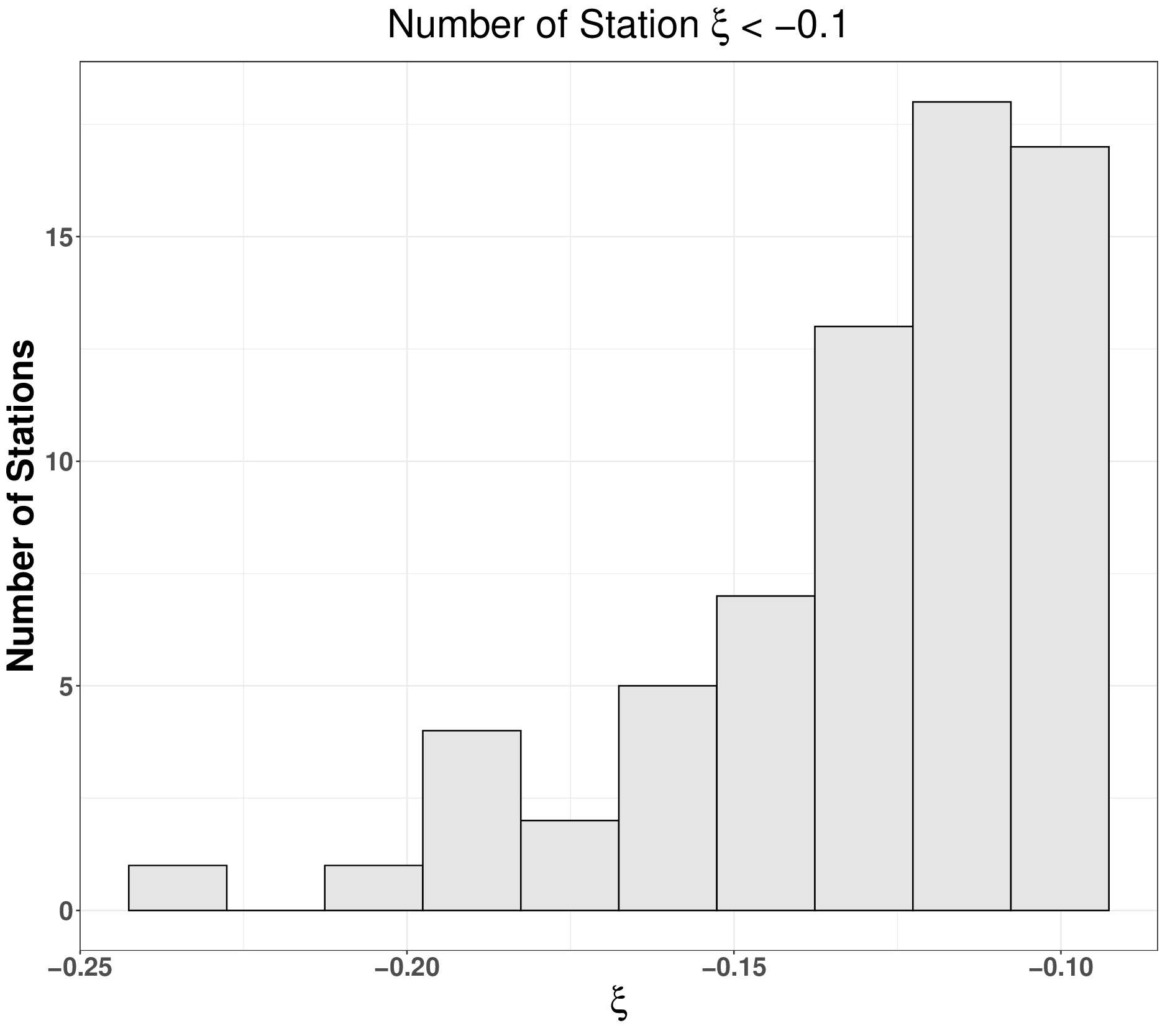}
        \caption{}
    \end{subfigure}
    \hfill
    \begin{subfigure}[b]{0.475\textwidth}
        \centering
        \includegraphics[width=\linewidth]{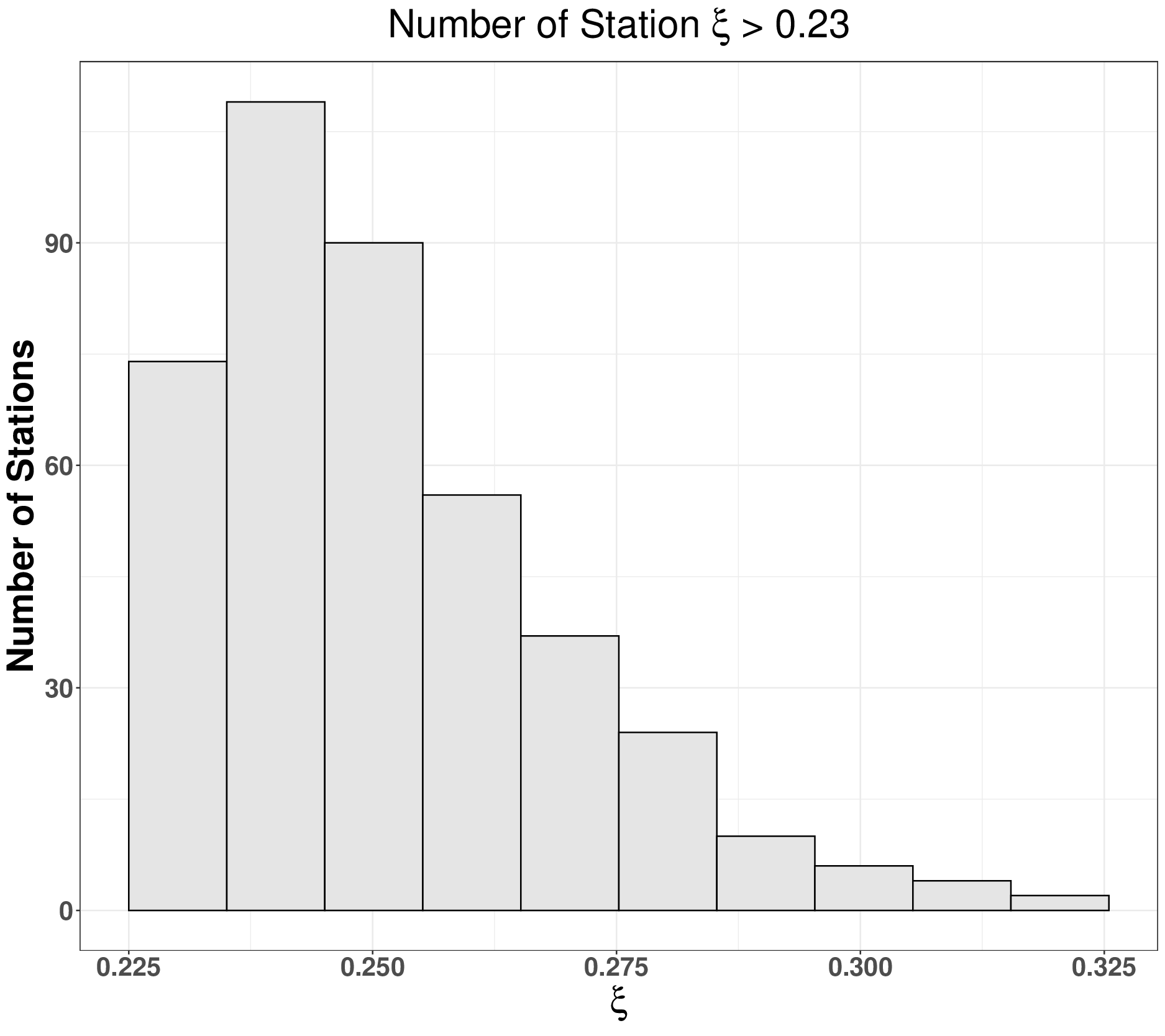}
        \caption{}
    \end{subfigure}
    \caption{The 480 stations whose shape parameter estimates fall outside the specified bounds after fitting the Beta prior. (a) Number of stations that estimated shape parameter less than $-0.1$; (b) Number of stations that estimated shape parameter greater than $0.23$.}
    \label{fig:BoundryFlags}
\end{figure}

Figure~\ref{fig:BoundryFlags} presents the categorized ranges for the 480 stations whose shape parameter estimates fall outside the specified bounds after applying the Beta prior.

\end{document}